\documentclass[journal]{journal}
\pdfoutput=1
\usepackage{multicol,lipsum}
\usepackage{booktabs}
\usepackage[normalem]{ulem}
\usepackage{longtable}
\useunder{\uline}{\ul}{}

\ifCLASSINFOpdf
	\usepackage[pdftex]{graphicx}
 
\else
\fi
\usepackage[cmex10]{amsmath}

\usepackage{mathtools}

\usepackage{float}

\usepackage[export]{adjustbox}

\usepackage{fixltx2e}
\usepackage{ulem}
\usepackage{subfigure}
\usepackage{xcolor}
\usepackage{setspace}

\usepackage[colorlinks,
            linkcolor=blue,
            anchorcolor=blue,
            citecolor=blue]{hyperref}

\usepackage{nameref}

\begin{document}
%
\title{Newtonian Mechanics Based Transient Stability PART IV: Equivalent Machine}
%
%
%

\author{{Songyan Wang,
        Jilai Yu,  
				Aoife Foley,
        Jingrui Zhang
        }
        
}
%
%

\markboth{Journal of \LaTeX\ Class Files,~Vol.~6, No.~1, January~2007}%
{Shell \MakeLowercase{\textit{et al.}}: Bare Demo of IEEEtran.cls for Journals}
%



\maketitle
\thispagestyle{empty}
\begin{abstract}
This paper analyzes the mechanisms of the equivalent machine and also its advantages in TSA. Based on the two group separations, an equivalent machine is modeled through the equivalence of the motions of all machines inside each group.
This ``motion equivalence” fully ensures the modeling of the two-machine system and the corresponding Newtonian energy conversion. Against this background, the original system becomes the equivalent system. It is clarified that the equivalent machine strictly follows the machine paradigms. These strict followings bring the two advantages in the equivalent-machine based TSA: (i) the stability of the equivalent machine is characterized precisely, and (ii) the equivalent-machine trajectory variance is depicted clearly. The two advantages are fully reflected in the precise definitions of the equivalent-machine based transient stability concepts. In particular, the equivalent machine swing is clearly depicted through the EDSP or EDLP of the machine, and the critical stability of the equivalent system is strictly defined as the critical stability of the equivalent machine. Simulation results show that the effectiveness of the equivalent-machine in TSA.

\end{abstract}

%

\IEEEpeerreviewmaketitle
\begin{small}

\begin{tabular}{lllll}
    &            &               &                  &                                \\
  \multicolumn{5}{c}{\textbf{Nomenclature}}                                                \\

  COI    & \multicolumn{1}{c}{}  & \multicolumn{3}{l}{Center of inertia}                   \\
  DLP    &                       & \multicolumn{3}{l}{Dynamic liberation point}            \\
  DSP    &                       & \multicolumn{3}{l}{Dynamic stationary point}            \\
  EAC    &                       & \multicolumn{3}{l}{Equal area criterion}                \\
  MPP    &                       & \multicolumn{3}{l}{Maximum potential energy point}      \\
  NEC    &                       & \multicolumn{3}{l}{Newtonian energy conversion}         \\
  TSA    &                       & \multicolumn{3}{l}{Transient stability assessment}      \\
  GTE    &                       & \multicolumn{3}{l}{Global total transient energy}           \\
  GMPP    &                       & \multicolumn{3}{l}{Global MPP}           \\
  IMKE    &                       & \multicolumn{3}{l}{Individual-machine kinetic energy}           \\ 
  IMPE    &                       & \multicolumn{3}{l}{Individual-machine potential energy}           \\ 
  IMTE    &                       & \multicolumn{3}{l}{Individual-machine total transient energy}           \\ 
  IMTR    &                       & \multicolumn{3}{l}{Individual-machine trajectory}           \\ 
  IVCS    &                       & \multicolumn{3}{l}{Individual-machine-virtual-COI-machine system}           \\ 
  IEEAC    &                       & \multicolumn{3}{l}{Integrated extended EAC}           \\  
  IMEAC    &                       & \multicolumn{3}{l}{Individual-machine EAC}           \\  

\end{tabular}
\end{small}

\section{Introduction} \label{section_I}

%
%
%
%
\subsection{LITERATURE REVIEW} \label{section_IA}
\raggedbottom
\IEEEPARstart{T}{he} global monitoring of the original system trajectory leads to the energy superimposition. This energy superimposition directly causes the superimposed machine to become a pseudo machine without equation of motion. Against this background, the pseudo superimposed machine completely violates all the machine paradigms.
These violations are also reflected in the defects of the definitions of the superimposed-machine based transient stability concepts \cite{1}-\cite{3}. This also inspires the global analysts to rethink about the dominant role of the two-machine modeling. Different from the global monitoring of the original system trajectory, Fouad conjectured that the original system can be separated as two groups, and all machines inside each group are aggregated as an equivalent machine \cite{4}, \cite{5}.
Following this equivalent-machine perspective, the system operator actually monitors the two equivalent-machine trajectories rather than the original system trajectory in TSA. Against this background, the original system becomes the equivalent system with two equivalent machines. The two-machine system is modeled successfully, and NEC (EAC) is established in this equivalent system. Further, the residual SMKE problem that occurs in the superimposed-machine is completely solved through this machine equivalence. Based on this equivalent-machine thinking, the famous IEEAC method was developed \cite{6}.
The equivalent machine shows both efficiency and precision compared with the superimposed machine based methods. It is also a milestone in the history of the transient stability analysis.
\par Compared with the ``energy superimposition”, the ``motion equivalence” can be seen as the keyword of the equivalent-machine. Based on the machine paradigms in the previous papers \cite{1}-\cite{3} and also the definitions of the equivalent machine, two questions can be emerged as follows:
\\ (i) Could the transient characteristics of the equivalent machine be explained from perspective of the individual-machine?
\\ (ii) Could the advantages of the equivalent machine be explained from the perspective of machine paradigms?
\par Obviously, answering the two questions may help readers take a deep insight into the mechanisms of the equivalent machine from the angle of ``motion equivalence”.

\subsection{SCOPE AND CONTRIBUTION OF THE PAPER} \label{section_IB}

Following the analysis in the previous two papers \cite{1}, \cite{2}, this paper focuses on the mechanisms of the equivalent-machine and its advantages in TSA through the ``motion equivalence” of the individual-machine.
\par Based on the group separations of the original system trajectory, the equivalent-machine trajectories are established. The relative motion of the two equivalent-machine trajectories are modeled through the two-machine system that is formed by the two equivalent machines, and in this way the NEC is finally established. Against this background, the equivalent machine becomes the ``one-and-only” machine in the equivalent system under a given group separation pattern. 
The transient characteristics of the equivalent machine are explained from the individual-machine perspective. It is clarified that the equivalent machine strictly follows all the machine paradigms. These strict followings bring the two advantages in TSA: (i) the equivalent-machine trajectory stability is characterized precisely (stability characterization advantage), and (ii) the equivalent-machine trajectory variance is depicted clearly at the equivalent-machine MPP (trajectory-depiction advantage).
After that, the two advantages of the equivalent machine are reflected in the definitions of the transient stability concepts. In particular, the one-and-only equivalent-machine swing is clearly depicted through its EMPP (reflection of trajectory-depiction advantage); The critical stability of the equivalent system is decided through the critical stability of the one-and-only equivalent machine (reflection of the two advantages). In the end of the paper, it is clarified that the equivalent-system stability is completely identical to the original-system stability. However, the equivalent-system severity might be different from the original-system severity because the inner-group-machine motion might be fierce under certain circumstances.
\par The contributions of this paper are summarized as follows:
\\ (i) The transient characteristics of the equivalent machine are systematically explained through the individual-machine perspective. This explains the mechanism of the equivalent-machine from an individual machine manner.
\\ (ii) The equivalent-machine transient stability is established based on the machine paradigms. This provides a precise modeling and stability characterization for the equivalent machine.
\\ (iii) All the transient stability concepts can be defined strictly through equivalent machine. This further validates the reasonability of the machine paradigms.
\par The reminder of the paper is organized as follows. In Section \ref{section_II}, the mechanisms of the equivalent-machine are analyzed. In Section \ref{section_III}, the characteristics of the equivalent machine are explained from the individual-machine perspective.
In Section \ref{section_IV}, the equivalent-machine based system stability is given through the strict followings of the machine paradigms. In Section \ref{section_V}, the advantages of the individual-machine based transient stability concepts are analyzed.
In Section \ref{section_VI}, simulation cases show the effectiveness of the equivalent-machine in TSA. In Section \ref{section_VII}, the relationship between the original-system and the equivalent-system is analyzed. In Section \ref{section_VIII}, detailed analysis about the mirror system is given. Conclusions are given in Section \ref{section_IX}.
\par In this paper, because the individual-machine and the equivalent-machine are defined in the COI-SYS and COI-NCR, respectively, the transformation between the two motion references is solved by using the mirror system. In addition, stability evaluations in this paper are mainly depicted by EAC rather than the transient energy.

\section{MECHANISMS OF THE EQUIVALENT-MACHINE} \label{section_II}
\subsection{EQUIVALENT MACHINE MONITORING}  \label{section_IIA}

The equivalent machine monitoring comprises two steps, i.e., the group formation and machine equivalence.
\\ \textit{Group formation}: The original system trajectory is first separated into two groups. A tutorial example about group separation is given in Fig. \ref{fig3}.
The original system trajectory is given in the synchronous reference. All machines in the system are separated as two groups, i.e, Group-CR ($\Omega_{\text{CR}}$) and Group-NCR ($\Omega_{\text{NCR}}$).
\par 
\begin{figure}[H]
  \centering
  \includegraphics[width=0.42\textwidth,center]{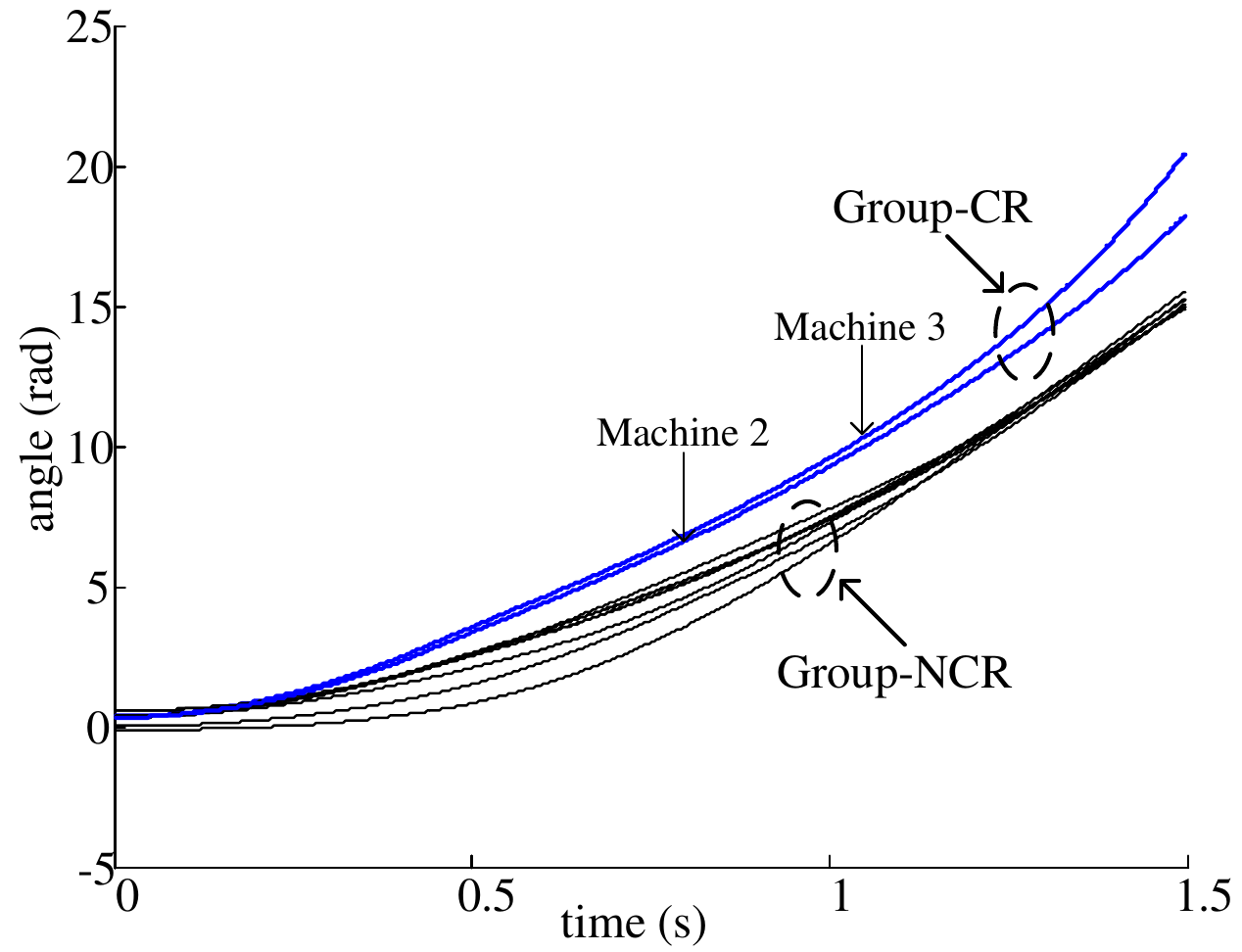}
  \caption{Group separation pattern in the synchronous reference [TS-1, bus-4, 0.447 s].} 
  \label{fig1}  
\end{figure}
\vspace*{-0.5em}
For the case in Fig. \ref{fig1}, At first glance, a natural group separation pattern should be Machines 2 and 3 forming Group-CR because the two machines are severely disturbed after fault clearing. However, theoretically, numerous group separation patterns can be formed in a multi-machine system.
\par Among all these possible patterns, the pattern with the lowest stability margin will be finally set as the dominant pattern in order to represent the original system. Note that all the equivalent-machine analysis is this paper given under the dominant pattern. Detailed analysis about the dominant pattern is given in Section \ref{section_IVB}.
\\ \textit{Motion equivalence}: the equivalent machine of each group is denoted as
\begin{equation}
  \label{equ1}
  \centering
  \begin{aligned}
    &\left\{\begin{array}{l}
    { \delta _ { \mathrm { CR } } = \frac { \sum _ { i \in \Omega _ { \mathrm { CR } } } M _ { i } \delta _ { i } } { M _ { \mathrm { CR } } } } \\
    \\
    { \omega _ { \mathrm { CR } } = \frac { \sum _ { i \in \Omega _ { \mathrm { CR } } } M _ { i } \omega _ { i } } { M _ { \mathrm { CR } } } } \\
    \\
    { P _ { \mathrm { CR } } = \sum _ { i \in \Omega _ { \mathrm { CR } } } ( P _ { m i } - P _ { e i } ) }
    \end{array}\right.\\
    & \left\{\begin{array}{l}
    \delta_{\mathrm{NCR}}=\frac{\sum_{j \in \Omega_{\mathrm{NCR}}} M_{j} \delta_{j}}{M_{\mathrm{NCR}}} \\
    \\
    \omega_{\mathrm{NCR}}=\frac{\sum_{j \in \Omega_{\mathrm{NCR}}} M_{j} \omega_{j}}{M_{\mathrm{NCR}}} \\
    \\
    P_{\mathrm{NCR}}=\sum_{j \in \Omega_{\mathrm{NCR}}}\left(P_{m j}-P_{e j}\right)
    \end{array}\right.
  \end{aligned} 
\end{equation}
where
\begin{spacing}{1.5}
  \noindent$M_{\mathrm{CR}}=\sum_{i \in \Omega_{\mathrm{CR}}} M_{i}$\\
  $M_{\mathrm{NCR}}=\sum_{j \in \Omega_{\mathrm{NCR}}} M_{j}$
\end{spacing}
Following Eq. (\ref{equ1}), the equation of motion of each equivalent Machine, i.e., Machine-CR and Machine-NCR are given as
\begin{equation}
  \label{equ2}
  \left\{\begin{array} { l } 
    { \frac { d \delta _ { \mathrm { CR } } } { d t } = \omega _ { \mathrm { CR } } }  \\
    \\
    { M _ { \mathrm { CR } } \frac { d \omega _ { \mathrm { CR } } } { d t } = P _ { \mathrm { CR } } }
    \end{array} \quad \left\{\begin{array}{l}
    \frac { d \delta _ { \mathrm { NCR } } } { d t } = \omega _ { \mathrm { NCR } }\\
    \\ 
    M_{\mathrm{NCR}} \frac{d \omega_{\mathrm{NCR}}}{d t}=P_{\mathrm{NCR}}
    \end{array}\right.\right.
\end{equation}
where
\begin{spacing}{1.5}
\noindent$P_{\mathrm{CR}}=\sum_{i \in \Omega_{\mathrm{CR}}}\left(P_{m i}-P_{e i}\right)$\\
$P_{\mathrm{NCR}}=\sum_{j \in \Omega_{\mathrm{NCR}}}\left(P_{m j}-P_{e j}\right)$
\end{spacing}
In Eq. (\ref{equ4}), the motion of each equivalent machine can be seen as the ``motion equivalence” of all machines in the group.
\par In the COI-NCR reference, the equivalent-machine trajectory (EMTR) is denoted as
\begin{equation}
  \label{equ3}
  \delta_{\mathrm{CR}\mbox{-}\mathrm{NCR}}=\delta_{\mathrm{CR}}-\delta_{\mathrm{NCR}}
\end{equation}
\par Based on Eq. (\ref{equ3}), the characteristics of the equivalent system trajectory are given as below
\vspace*{0.5em}
\\
(i) The equivalent Machine-NCR is set as the RM.
\\
(ii) Using the Machine-NCR as the motion reference, he equivalent system trajectory is formed by the ``one-and-only” EMTR of Machine-CR.
\vspace*{0.5em}
\par (i) and (ii) indicate the following
\vspace*{0.5em}
\par \textit{The equivalent system trajectory is formed by the ``one-and-only” EMTR of Machine-CR in the COI-NCR reference under the given group separation pattern.}
\vspace*{0.5em}
\par That is, the equivalent-machine stability is identical to the equivalent-system stability. The ``equivalent machine” is the same as the ``equivalent system”.
\par A tutorial example is given below to demonstrate the equivalent system trajectory. The equivalent system trajectory in the synchronous reference and that in the COI-NCR reference are shown in Figs. \ref{fig2} (a) and (b), respectively.
\begin{figure} [H]
  \centering 
  \subfigure[]{%
  \label{fig1a}
    \includegraphics[width=0.37\textwidth]{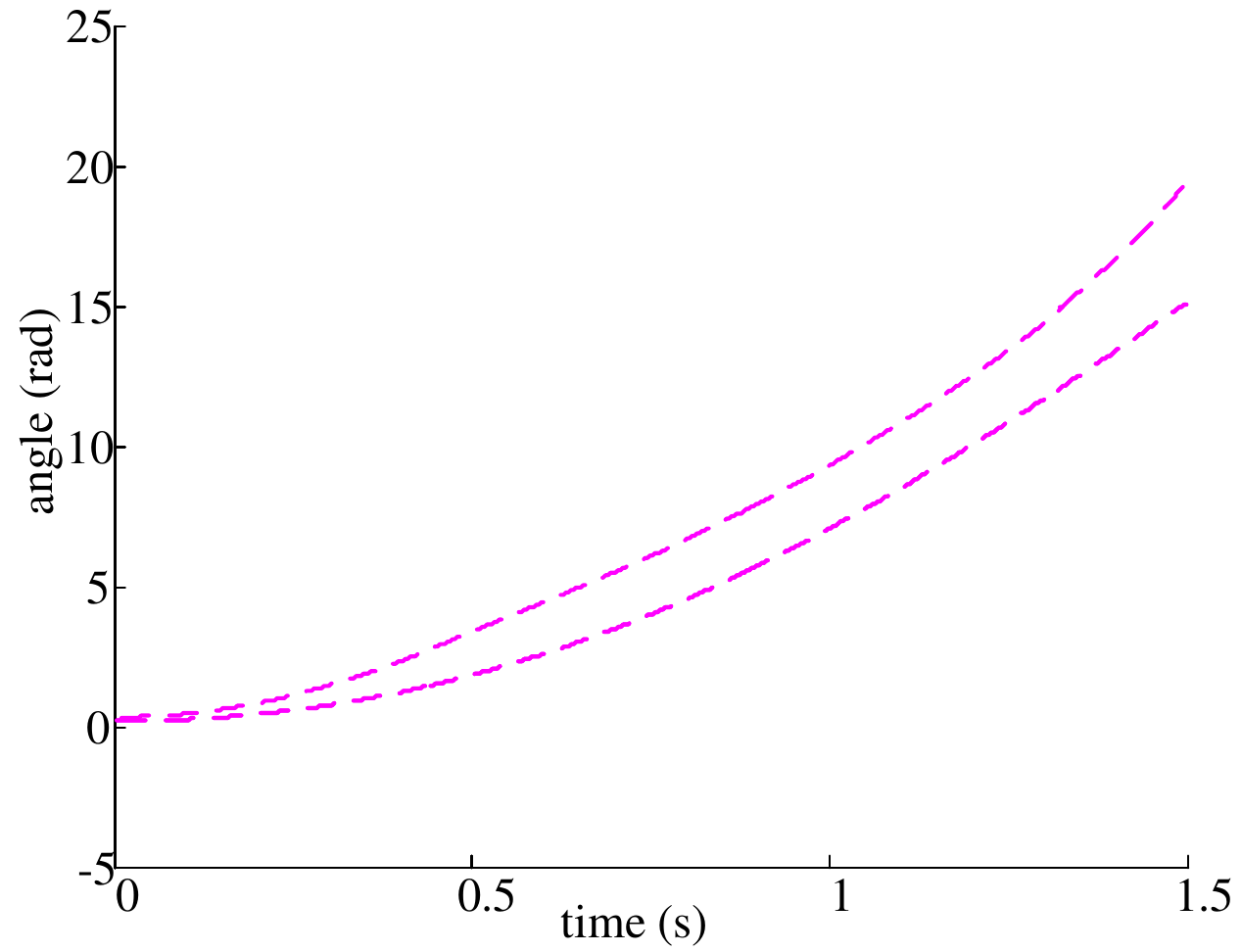}}%
\end{figure} 
\addtocounter{figure}{-1}
\vspace*{-2em}       
\begin{figure} [H]
  \addtocounter{figure}{1}      
  \centering 
  \subfigure[]{%
    \label{fig1b}
    \includegraphics[width=0.37\textwidth]{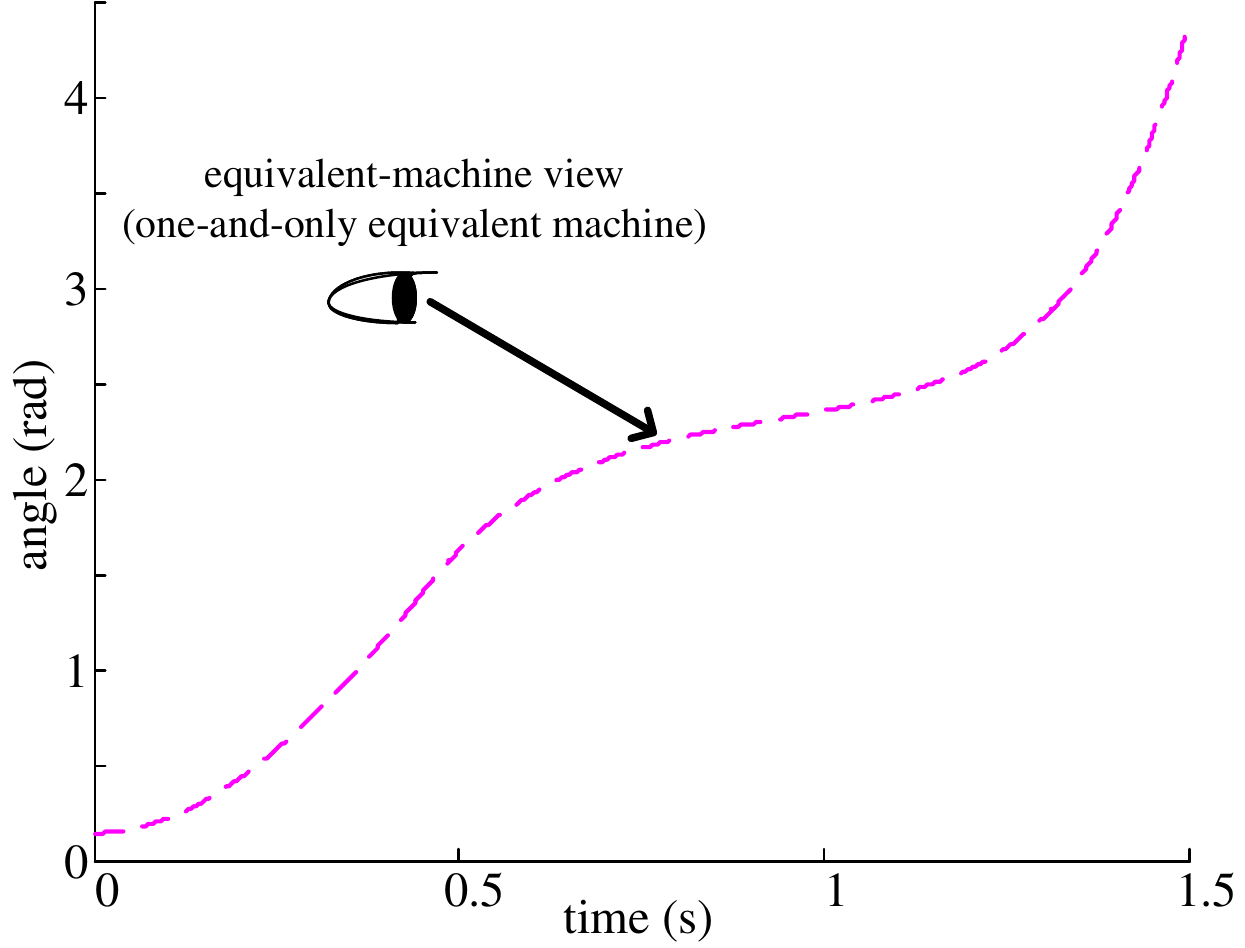}}%
  \caption{Equivalent system trajectory [TS-1, bus-4, 0.447 s]. (a) EMTR in the synchronous reference. (b) EMTR in the COI-NCR reference.}%
  \label{fig2}
\end{figure}
\vspace*{-0.5em}
From Fig. \ref{fig2}, through the equivalence of the original system trajectory, the equivalent system trajectory is formed by the one-and-only EMTR of the equivalent Machine-CR in the COI-NCR reference under a certain group separation pattern. Therefore, the equivalent system is different from the original system that is formed by multiple individual machines without any equivalence.

\subsection{CR-NCR SYSTEM MODELING} \label{section_IIB}
\begin{figure}[H]
  \centering
  \includegraphics[width=0.45\textwidth,center]{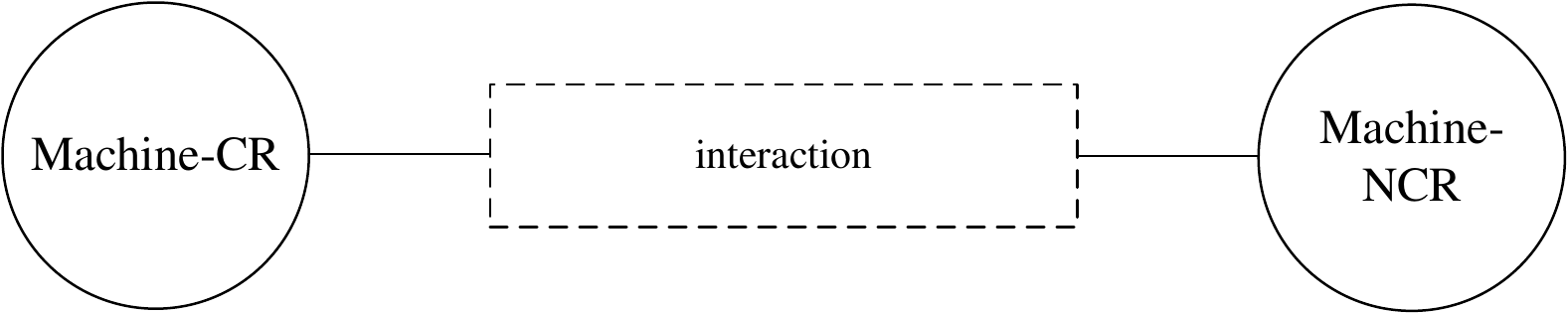}
  \caption{Formation of the CR-NCR system.} 
  \label{fig3}  
\end{figure}

Based on the equivalent-machine monitoring, the variance of the $\delta_{\text{CR-NCR}}$ is modeled through the corresponding two-machine system, i.e., the Machine-CR-Machine-NCR system (CR-NCR system). The formation of the CR-NCR system is shown in Fig. \ref{fig3}.
Based on Eq. (\ref{equ1}), the modeling of the CR-NCR system is depicted as
\begin{equation}
  \label{equ4}
  \left\{\begin{array}{l}
    \frac{d \delta_{\mathrm{CR}\mbox{-}\mathrm{NCR}}}{d t}=\omega_{\mathrm{CR}\mbox{-}\mathrm{NCR}} \\\\
    M_{\mathrm{CR}} \frac{d \omega_{\mathrm{CR}\mbox{-}\mathrm{NCR}}}{d t}=f_{\mathrm{CR}\mbox{-}\mathrm{NCR}}
    \end{array}\right.
\end{equation}
where
\begin{spacing}{1.5}
  \noindent$\delta_{\mathrm{CR}\mbox{-}\mathrm{NCR}}=\delta_{\mathrm{CR}}-\delta_{\mathrm{NCR}}$\\
  $\omega_{\mathrm{CR}\mbox{-}\mathrm{NCR}}=\omega_{\mathrm{CR}}-\omega_{\mathrm{NCR}}$\\
  $f_{\mathrm{CR}\mbox{-}\mathrm{NCR}}=P_{\mathrm{CR}}-\frac{M_{\mathrm{CR}}}{M_{\mathrm{NCR}}} P_{\mathrm{NCR}}$
\end{spacing}
From Eq. (\ref{equ4}), $\delta_{\mathrm{CR}\mbox{-}\mathrm{NCR}}$ is completely modeled through the relative motion in the CR-NCR system.
\par The equivalent-machine DLP (EDLP) is denoted as 
\begin{equation}
  \label{equ5}
  f_{\mathrm{CR}\mbox{-}\mathrm{NCR}}=0
\end{equation}
\par In Eq. (\ref{equ5}), the EDLP of Machine-CR depicts the point where the equivalent machine becomes unstable.

\subsection{EQUIVALENT MACHINE TRANSIENT ENERGY CONVERSION} \label{section_IIC}
The EMTE is defined in a typical Newtonian energy manner. The EMTE is defined as
\begin{equation}
  \label{equ6}
  V_{\mathrm{CR}\mbox{-}\mathrm{NCR}}=V_{KE\mathrm{CR}\mbox{-}\mathrm{NCR}}+V_{PE\mathrm{CR}\mbox{-}\mathrm{NCR}}
\end{equation}
where
\begin{spacing}{2}
  \noindent$V_{K E \mathrm{C R}\mbox{-}\mathrm{NCR}}=\frac{1}{2} M_{\mathrm{CR}} \omega_{\mathrm{CR}\mbox{-}\mathrm{NCR}}^{2}$\\
  $V_{P E \mathrm{CR}\mbox{-}\mathrm{NCR}}=\int_{\delta_{\mathrm{CR}\mbox{-}\mathrm{NCR}}^{s}}^{\delta_{\mathrm{CR}\mbox{-}\mathrm{NCR}}}\left[-f_{\mathrm{CR}\mbox{-}\mathrm{NCR}}^{(P F)}\right] d \delta_{\mathrm{CR}\mbox{-}\mathrm{NCR}}$
\end{spacing}
In Eq. (\ref{equ6}), the conversion between EMKE and EMPE is used to measure the stability of the CR-NCR system.
\par The residual EMKE of at its corresponding EMPP is denoted as
\begin{equation}
  \label{equ7}
  \begin{split}
    V_{K E \mathrm{C R}\mbox{-}\mathrm{NCR}}^{R E}&=V_{K E\mathrm{ C R}\mbox{-}\mathrm{NCR}}^{c}-\Delta V_{P E \mathrm{C R}\mbox{-}\mathrm{NCR}} \\ 
    &=A_{A C C \mathrm{C R}\mbox{-}\mathrm{N C R}}-A_{D E C \mathrm{C R}\mbox{-}\mathrm{NCR}}
  \end{split}
\end{equation}
where
\begin{spacing}{2}
  \noindent$V_{K E \mathrm{C R}\mbox{-}\mathrm{NCR}}^{c}=\frac{1}{2} M_{\mathrm{CR}} \omega_{\mathrm{CR}\mbox{-}\mathrm{NCR}}^{c 2}=A_{A C C \mathrm{C R}\mbox{-}\mathrm{NCR}}$ \\
  $\Delta V_{P E C\mathrm{CR}\mbox{-}\mathrm{NCR}}=\int_{\delta_{\mathrm{CR}\mbox{-}\mathrm{NCR}}^{s}}^{\delta_{\mathrm{CR}\mbox{-}\mathrm{NCR}}^{EMPP}}\left[-f_{\mathrm{CR}\mbox{-}\mathrm{NCR}}^{(P F)}\right] d \delta_{\mathrm{CR}\mbox{-}\mathrm{NCR}}$\\
  $-\int_{\delta_{\mathrm{CR}\mbox{-}\mathrm{NCR}}^{s}}^{\delta_{\mathrm{CR}\mbox{-}\mathrm{NCR}}^{c}}\left[-f_{\mathrm{CR}\mbox{-}\mathrm{NCR}}^{(P F)}\right] d \delta_{\mathrm{CR}\mbox{-}\mathrm{NCR}}=A_{DEC\mathrm{CR}\mbox{-}\mathrm{NCR}}$
\end{spacing}
In Eq. (\ref{equ7}), note that the equivalent-machine transient energy conversion is identical to the EMEAC. This proof is simple because it is similar to the individual-machine case as analyzed in Ref. \cite{1}.
\par The stability characterizations of the equivalent machine are summarized as below.
\\
(i) From the perspective of transient energy conversion, Machine-CR is evaluated to go unstable if the residual EMKE occurs at its EMPP.\\
(ii) From the perspective of EAC, the equivalent machine is evaluated to go unstable if the acceleration area is larger than the deceleration area.
\vspace*{0.5em}
\par
\begin{figure}[H]
  \centering
  \includegraphics[width=0.45\textwidth,center]{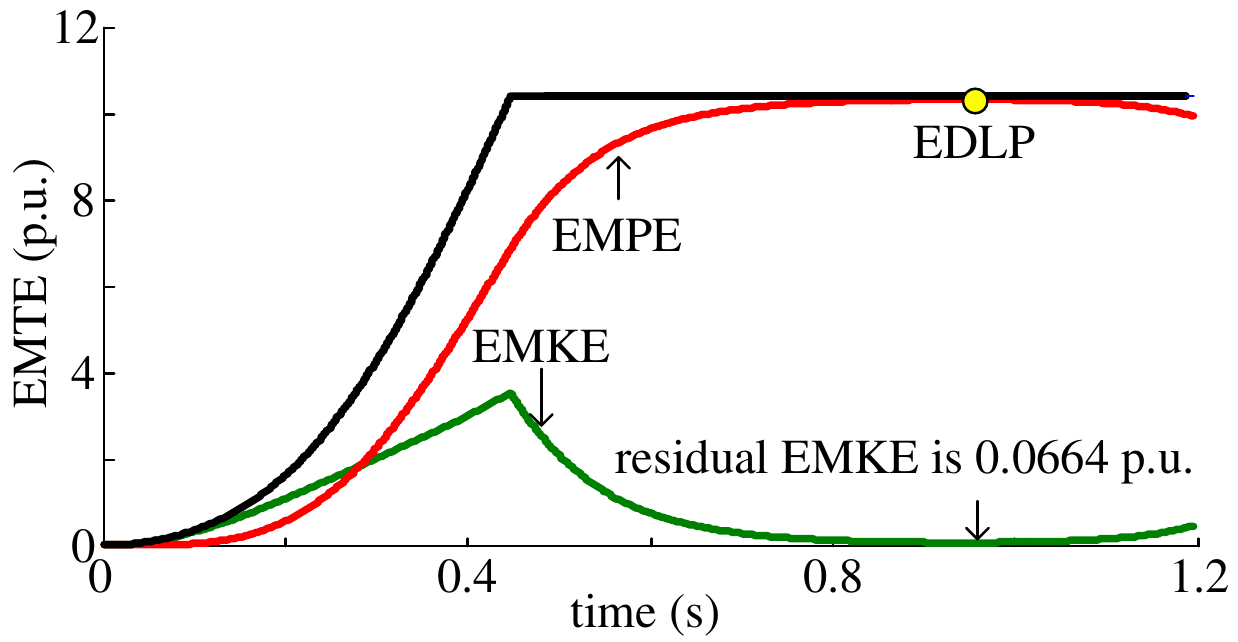}
  \caption{Equivalent machine transient energy conversion [TS-1, bus-4, 0.447s].} 
  \label{fig4}  
\end{figure}

\begin{figure}[H]
  \centering
  \includegraphics[width=0.42\textwidth,center]{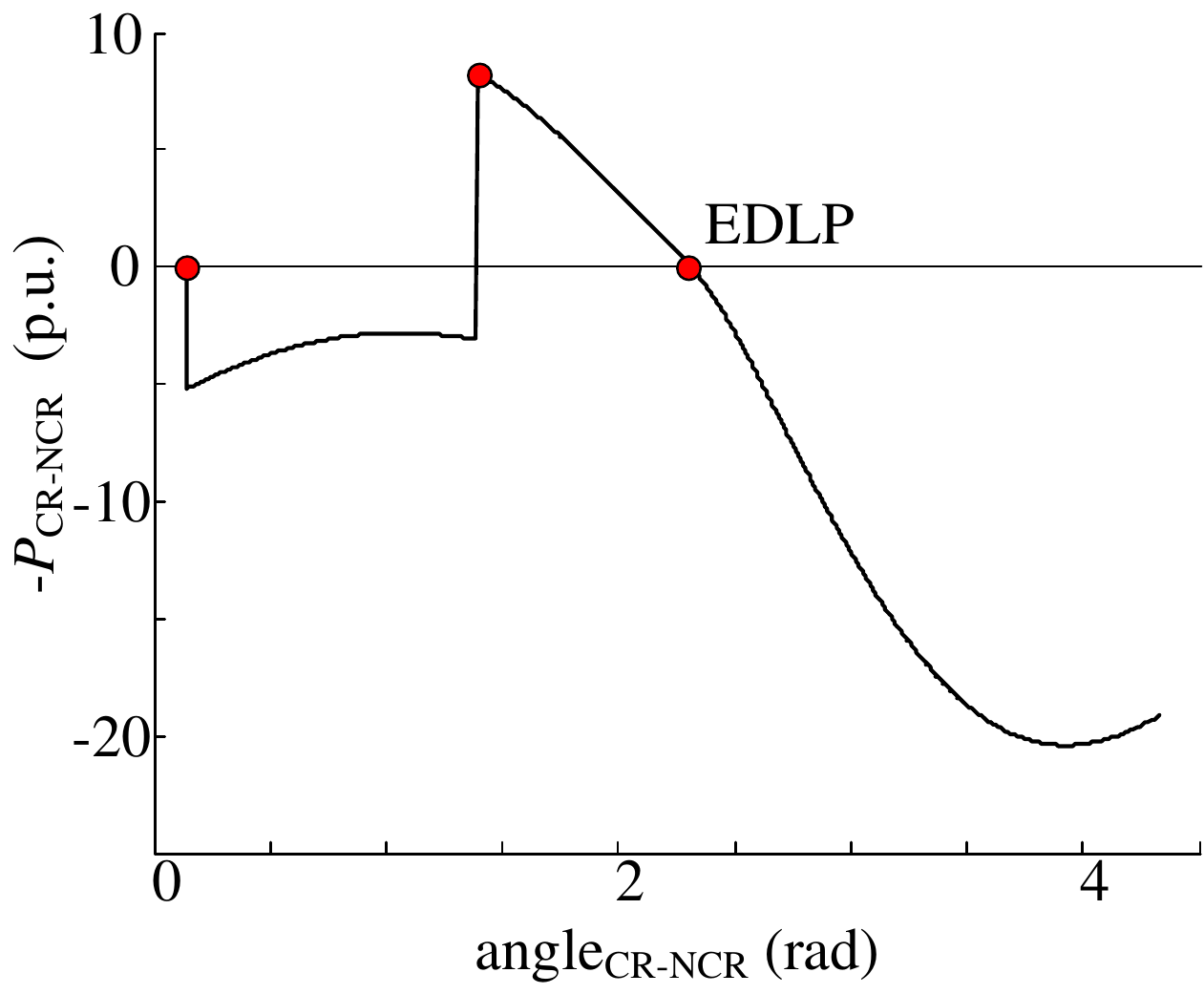}
  \caption{EMEAC [TS-1, bus-4, 0.447s].} 
  \label{fig5}  
\end{figure}

 For the case as in Fig. \ref{fig1}, the equivalent-machine transient energy conversion is shown in Fig. \ref{fig4}. The EMEAC is shown in Fig. \ref{fig5}.
Note again that the equivalent-machine transient energy conversion is completely identical to the EMEAC. The two are just the different expressions of the transient energy conversion in the \textit{t-V} space and $\delta\mbox{-}f$ space, respectively.
\vspace*{-1em}
From Figs. \ref{fig4} and \ref{fig5}, the equivalent machine is the ``one-and-only” machine in TSA because this machine is defined based on the motion equivalence of all machines in the two groups. The equivalent machine analysts only monitor the EMEAC.
\par The characteristics of the EMEAC are given as below.\\
(Characteristic-I) The equivalent trajectory occurs among IMTRs of machines in $\Omega_{\mathrm{CR}}$ at a certain time point.\\
(Characteristic-II) EDSP occurs among the IDSPs of the machines in $\Omega_{\mathrm{CR}}$ along time horizon.\\
(Characteristic-III) EDLP occurs among the IDLPs of all machines in $\Omega_{\mathrm{CR}}$ along time horizon.
\vspace*{0.5em}
\par Because the equivalent machine is defined based on the motion equivalence of all machines in the two groups, in the following section, all the characteristics above will be explained in an individual-machine manner. Detailed analysis is given in Section \ref{section_III}. This may help readers deeply understand the ``motion equivalence” of the superimposed machine.
\par Since the individual-machine and the equivalent machine are defined in the different motion references (COI-SYS reference and the COI-NCR reference), in the following section, the ``mirror system” of the CR-NCR system, i.e., the CR-SYS system will be established first.
In this way both the equivalent machine and the individual-machine can be analyzed under the same COI-SYS reference. Note that the description in the following section is given in brief with only conclusions. Detailed analysis is provided in Section \ref{section_VIII}.

\section{EXPLANATION OF THE EQUIVALENT MACHINE FROM INDIVIDUAL-MACHINE PERSPECTIVE} \label{section_III}
\subsection{MIRROR SYSTEM} \label{section_IIIA}
Following the analysis as given in Section \ref{section_IIA}, the parameters of Machine-CR in the COI-SYS reference is denoted as
\begin{equation}
  \label{equ8}
  \left\{\begin{array}{l}
    \delta_{\mathrm{CR}\mbox{-}\mathrm{SYS}}=\frac{\sum_{i \in \Omega_{\mathrm{CR}}} M_{i} \delta_{i\mbox{-}\mathrm{SYS}}}{M_{\mathrm{CR}}} \\
    \\
    \omega_{\mathrm{CR}\mbox{-}\mathrm{SYS}}=\frac{\sum_{i \in \Omega_{\mathrm{CR}}} M_{i} \omega_{i\mbox{-}\mathrm{SYS}}}{M_{\mathrm{CR}}} \\
    \\
    f_{\mathrm {CR}\mbox{-}\mathrm{SYS}}=\sum_{i \in \Omega_{\mathrm{CR}}} f_{i\mathrm{-}\mathrm{SYS}}
    \end{array}\right.
\end{equation}
\par The parameters in Eq. (\ref{equ9}) can also be depicted as
\begin{equation}
  \label{equ9}
  \left\{\begin{array}{l}
    \delta_{\mathrm{CR}\mbox{-}\mathrm{SYS}}=\delta_{\mathrm{CR}}-\delta_{\mathrm{SYS}} \\
    \\
    \omega_{\mathrm{CR}\mbox{-}\mathrm{SYS}}=\omega_{\mathrm{CR}}-\omega_{\mathrm{SYS}} \\
    \\
    f_{\mathrm{CR}\mbox{-}\mathrm{SYS}}=P_{\mathrm{CR}}-\frac{M_{\mathrm{CR}}}{M_{\mathrm{SYS}}} P_{\mathrm{SYS}}
    \end{array}\right.
\end{equation}
\par Based on Eq. (\ref{equ9}), The equations of motions of Machine-CR in the COI-SYS reference can be depicted as\
\begin{equation}
  \label{equ10}
  \left\{\begin{array}{l}
    \frac{d \delta_{\mathrm{CR}\mbox{-}\mathrm{SYS}}}{d t}=\omega_{\mathrm{CR}\mbox{-}\mathrm{SYS}} \\
    \\
    M_{\mathrm{CR}} \frac{d \omega_{\mathrm{CR}\mbox{-}\mathrm{SYS}}}{d t}=f_{\mathrm{CR}\mbox{-}\mathrm{SYS}}
    \end{array}\right.
\end{equation}
\par
\begin{figure}[H]
  \centering
  \includegraphics[width=0.45\textwidth,center]{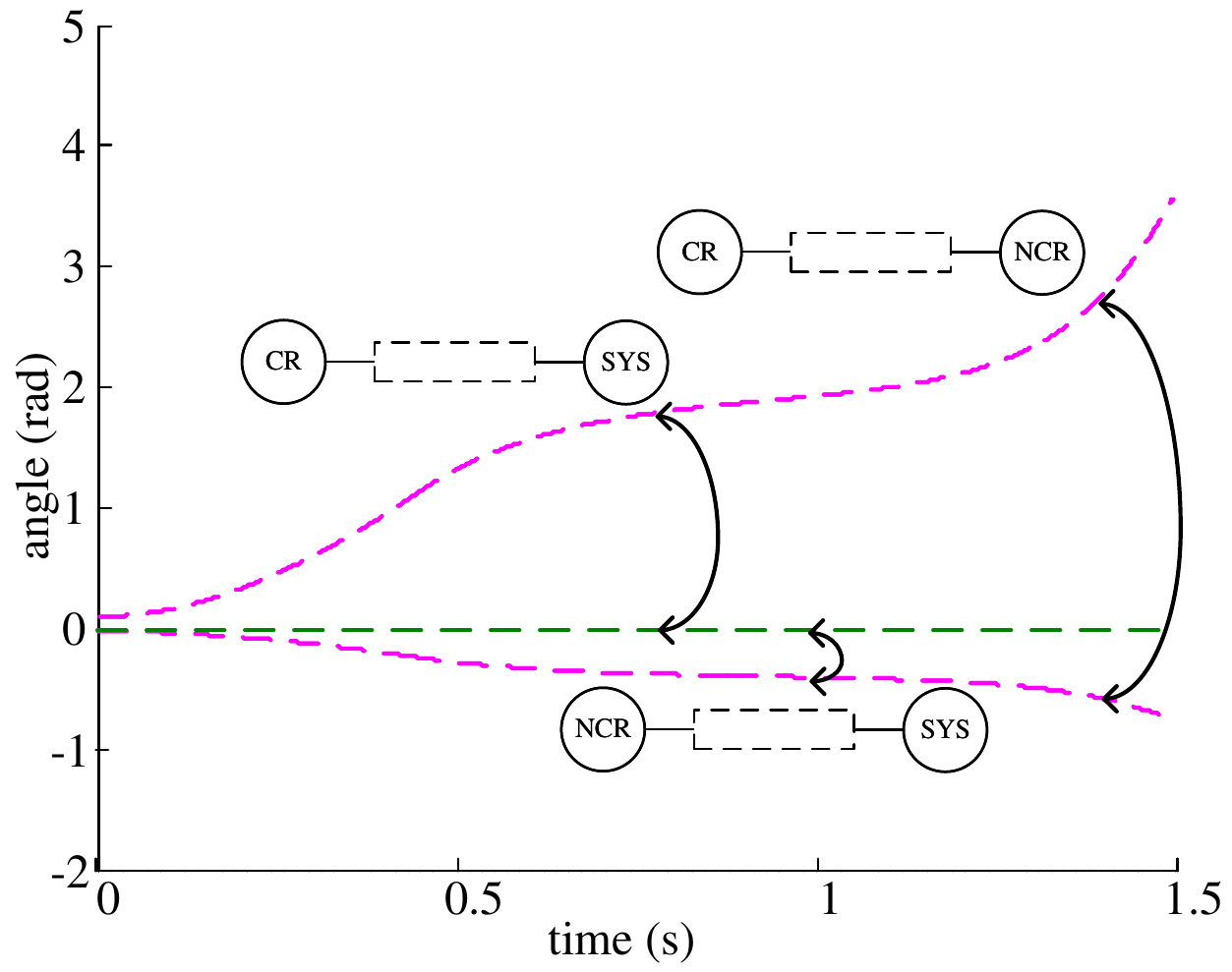}
  \caption{The CR-SYS system and CR-NCR system in the COI-SYS reference [TS-1, bus-4, 0.447 s].} 
  \label{fig6}  
\end{figure}

 From Eq. (\ref{equ10}), both Machine-CR and Machine-SYS form CR-SYS system. Demonstrations about the CR-SYS system and the CR-NCR system is shown in Fig. \ref{fig6}.
\vspace*{-0.5em}
\par The relationship between CR-NCR system and the CR-SYS system is depicted as
\begin{equation}
  \label{equ11}
  \left\{\begin{array}{l}
    \frac{d \delta_{\mathrm{CR}\mbox{-}\mathrm{NCR}}}{d t}=\frac{M_{\mathrm{SYS}}}{M_{\mathrm{NCR}}} \frac{d \delta_{\mathrm{CR}\mbox{-}\mathrm{SYS}}}{d t} \\
    \\
    \frac{d \omega_{\mathrm{CR}\mbox{-}\mathrm{NCR}}}{d t}=\frac{M_{\mathrm{SYS}}}{M_{\mathrm{NCR}}} \frac{d \omega_{\mathrm{CR}-\mathrm{SYS}}}{d t}
    \end{array}\right.
\end{equation}
\par From Eq. (11), one can obtain the following 
\vspace*{0.5em}
\\
(i) The motion of the CR-SYS system can be depicted as a constant ratio ($M_{\text{SYS}}$/$M_{\text{NCR}}$) of the motion of the CR-NCR system, respectively.
\\
(ii) The stability margins of the two systems are completely identical.
\vspace*{0.5em}
\par Following (i) to (iii), the CR-SYS system can be seen as the scale-down ``mirror system” of the CR-NCR system. Note that the NCR-SYS system can also be used as the mirror system of the CR-NCR system (detailed analysis is given in Section \ref{section_VIII}). 
\par In the following paper, all the stability analysis corresponding to the CR-NCR system will be replaced with the CR-SYS system. In this way both the equivalent machine and the individual-machine can be analyzed under the same COI-SYS reference.

\subsection{EQUIVALENT TRAJECTORY (CHARACTERISTIC-I)} \label{section_IIIB}
\noindent \textit{Explanation}: The EMTR can be seen as the equivalence of the IMTRs of all the stable machines in Group-CR.
\\
\textit{Analysis}: At a certain time oint \textit{t} along time horizon, assume the maximum angle and minimum angle occur at Machines \textit{m} and \textit{n}, respectively. Then the following holds
\begin{equation}
  \label{equ12}
  M_{\mathrm{CR}} \delta_{n\mbox{-}\mathrm{SYS}}<\sum_{i \in \Omega_{\mathrm{CR}}} M_{i} \delta_{i\mbox{-}\mathrm{SYS}}<M_{\mathrm{CR}} \delta_{m-\mathrm{SYS}}
\end{equation} 
\par Following Eq. (\ref{equ8}), Eq. (\ref{equ12}) can be further denoted as
\begin{equation}
  \label{equ13}
  \delta_{n\mbox{-}\mathrm{SYS}}<\delta_{\text {CR-SYS }}<\delta_{m\mbox{-}\mathrm{SYS}}
\end{equation} 
\par Eq. (\ref{equ13}) fully indicates that The EMTR lies among IMTRs of machines in $\Omega_{\mathrm{CR}}$ at any time point. This can be simply extended to the case of the equivalent velocity ($\omega_{\mathrm{CR}\mbox{-}\mathrm{SYS}}$).
\par The equivalent motion of Machine-CR (in the COI-SYS reference) is demonstrated as below. In this case $\Omega_{\mathrm{CR}}$=\{Machine 2, Machine 3\}. The angle and velocity of Machine-CR at 1.000 s is shown in Table \ref{table1}. Note that all the simulations below are given in the COI-SYS reference by using the mirror system.
\begin{figure} [H]
  \centering 
  \subfigure[]{%
  \label{fig7a}
    \includegraphics[width=0.4\textwidth]{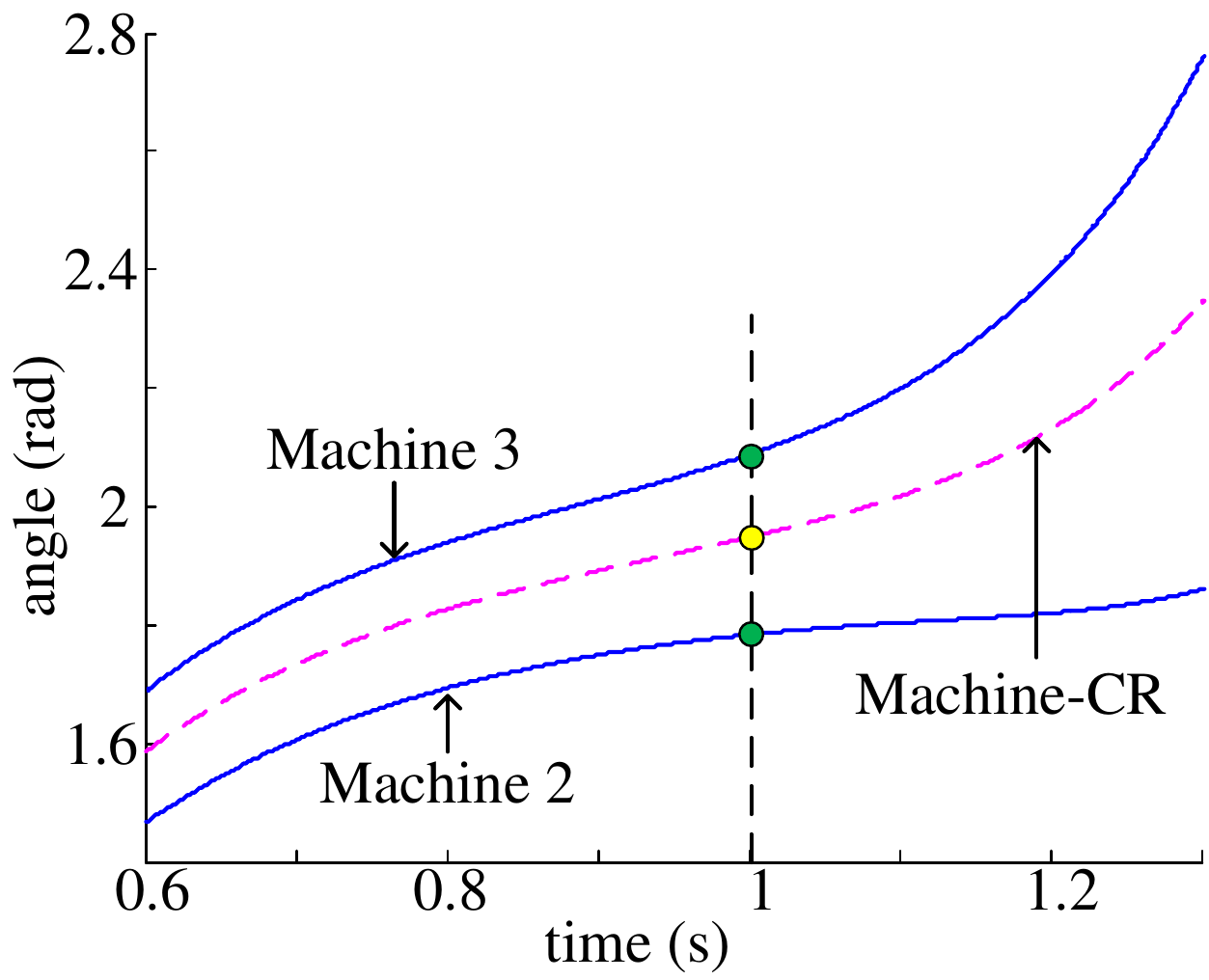}}%
\end{figure} 
\vspace*{-2em}
\addtocounter{figure}{-1}       
\begin{figure} [H]
  \addtocounter{figure}{1}      
  \centering 
  \subfigure[]{%
    \label{fig7b}
    \includegraphics[width=0.4\textwidth]{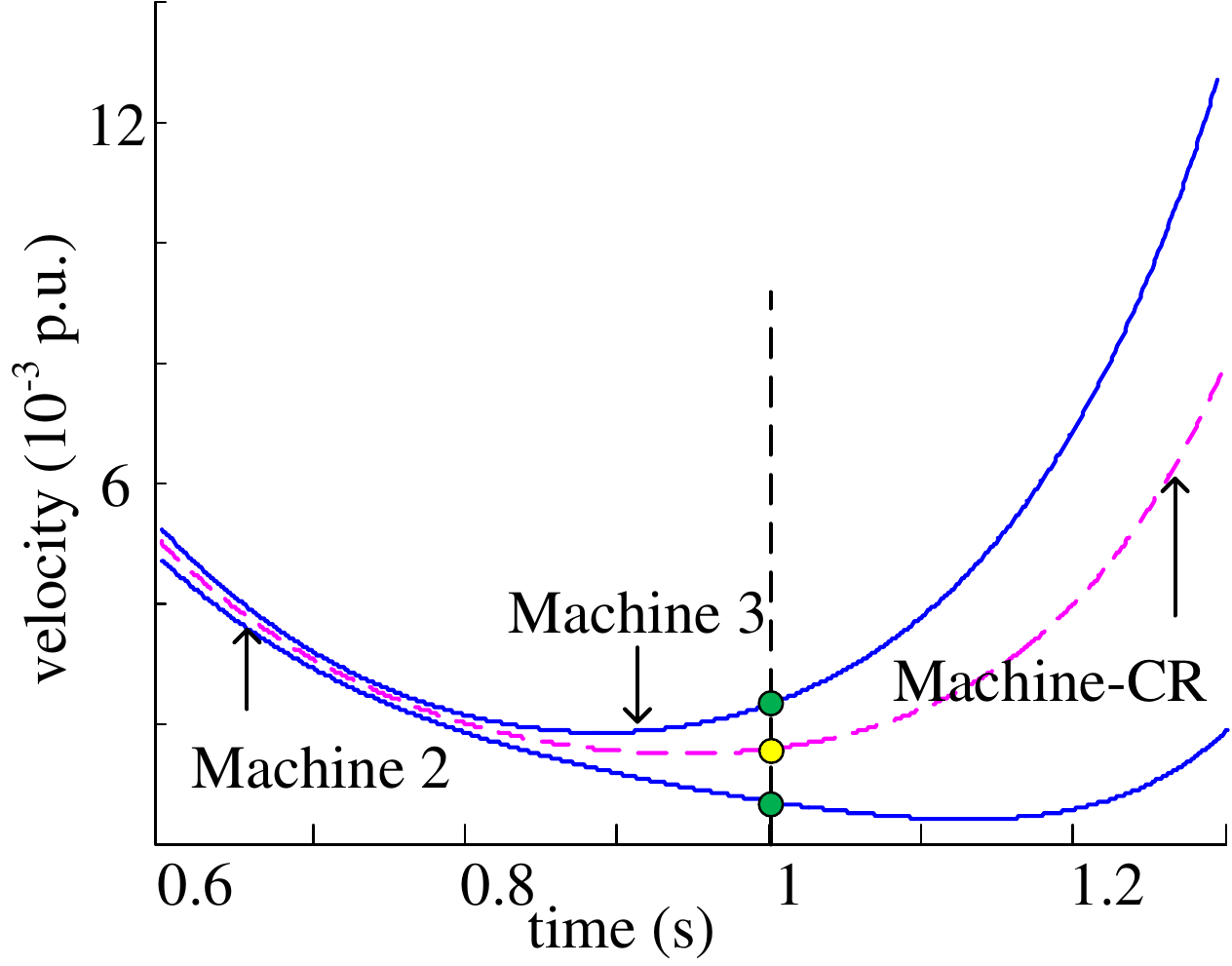}}%
  \caption{Equivalent motion of Machine-CR [TS-1, bus-4, 0.447 s].}%
  \label{fig7}
\end{figure}
\vspace*{-1em}
\begin{table}[H]
  \centering
  \caption{Angle and velocity of Machine-CR at 1.000 s}
  \begin{tabular}{@{}cccc@{}}
  \toprule
                  & Machine-CR & Machine 2 & Machine 3 \\ \midrule
  Angle (red)     & 1.9472     & 1.7829    & 2.0863    \\
  Velocity (p.u.) & 0.0016     & 0.0007    & 0.0023    \\
  Inertia (p.u.)  & 132.2      & 60.6      & 71.6      \\ \bottomrule
  \end{tabular}
  \label{table1}
\end{table}
\vspace*{-0.5em}
rom Fig. \ref{fig7} and Table \ref{table1}, the motion of Machine-CR can be seen as the equivalence of the motions of two physical machines in Group-CR.
In particular, following Eq. (\ref{equ9}), $\delta_{\mathrm{CR}\mbox{-}\mathrm{SYS}}$ is smaller than $\delta_{3\mbox{-}\mathrm{SYS}}$ yet larger than $\delta_{2\mbox{-}\mathrm{SYS}}$ at 1.000 s. The $\omega_{\mathrm{CR}\mbox{-}\mathrm{SYS}}$ also shows a similar equivalent characteristic.
\par From analysis above, Machine-CR can be seen as the special equivalent case of the individual-machine. In fact, the motion equivalence can also be reflected in the occurrence of the EDSP and EDLP. The analysis is given in the following sections.

\subsection{OCCURRENCE OF EDSP (CHARACTERISTIC-II)} \label{section_IIIC}
\noindent \textit{Explanation}: The EDSP can be seen as the equivalence of the IDSPs of all the stable machines in Group-CR.
\\
\textit{Analysis}: Once $\text{DSP}_{\text{CR-SYS}}$ occurs along time horizon, $\omega_{\text{CR-SYS}}$ will reach zero, and thus the following holds
\begin{equation}
  \label{equ14}
  \sum_{i \in \Omega_{\mathrm{CR}}} M_{i} \omega_{i\mbox{-}\mathrm{SYS}}=M_{\mathrm{CR}} \omega_{\mathrm{CR}\mbox{-}\mathrm{SYS}}=0
\end{equation}
\par Eq. (\ref{equ14}) indicates the following
\vspace*{0.5em}
\\
(i) $\omega_{\mathrm{CR}\mbox{-}\mathrm{SYS}}$ cannot reach zero if all $\omega_{i\mbox{-}\mathrm{SYS}}$ are positive.\\
(ii) $\omega_{\mathrm{CR}\mbox{-}\mathrm{SYS}}$ cannot reach zero if all $\omega_{i\mbox{-}\mathrm{SYS}}$ are negative.\\
(iii) $\omega_{\mathrm{CR}\mbox{-}\mathrm{SYS}}$ may reach zero only when some $\omega_{i\mbox{-}\mathrm{SYS}}$ become negative while the rest still remains positive.
\vspace*{0.5em}
\par From analysis above, $\text{DSP}_{\text{CR-SYS}}$ will occur among the IDSPs of the machines in $\Omega_{\mathrm{CR}}$ along time horizon. At the moment that $\text{DSP}_{\text{CR-SYS}}$ occurs,
some critical machines in $\Omega_{\mathrm{CR}}$ already inflect back in their second swings with negative $\omega_{i\mbox{-}\mathrm{SYS}}$, while the other critical machines in $\Omega_{\mathrm{CR}}$ are still moving in the first swing with positive $\omega_{i\mbox{-}\mathrm{SYS}}$.
\par The occurrence of $\text{EDSP}_{\text{CR-SYS}}$ is demonstrated in Fig. \ref{fig10}. In this case both Machines 2 and 3 maintain critical stable. The occurrence of the IDSP of each machine is also shown in the figure. The Kimbark curve of Machine-CR in the COI-SYS reference is shown in Fig. \ref{fig11}.
The velocity of the each critical machine at $\text{DSP}_{\text{CR-SYS}}$ are shown in Table \ref{table2}.
\begin{figure}[H]
  \centering
  \includegraphics[width=0.4\textwidth,center]{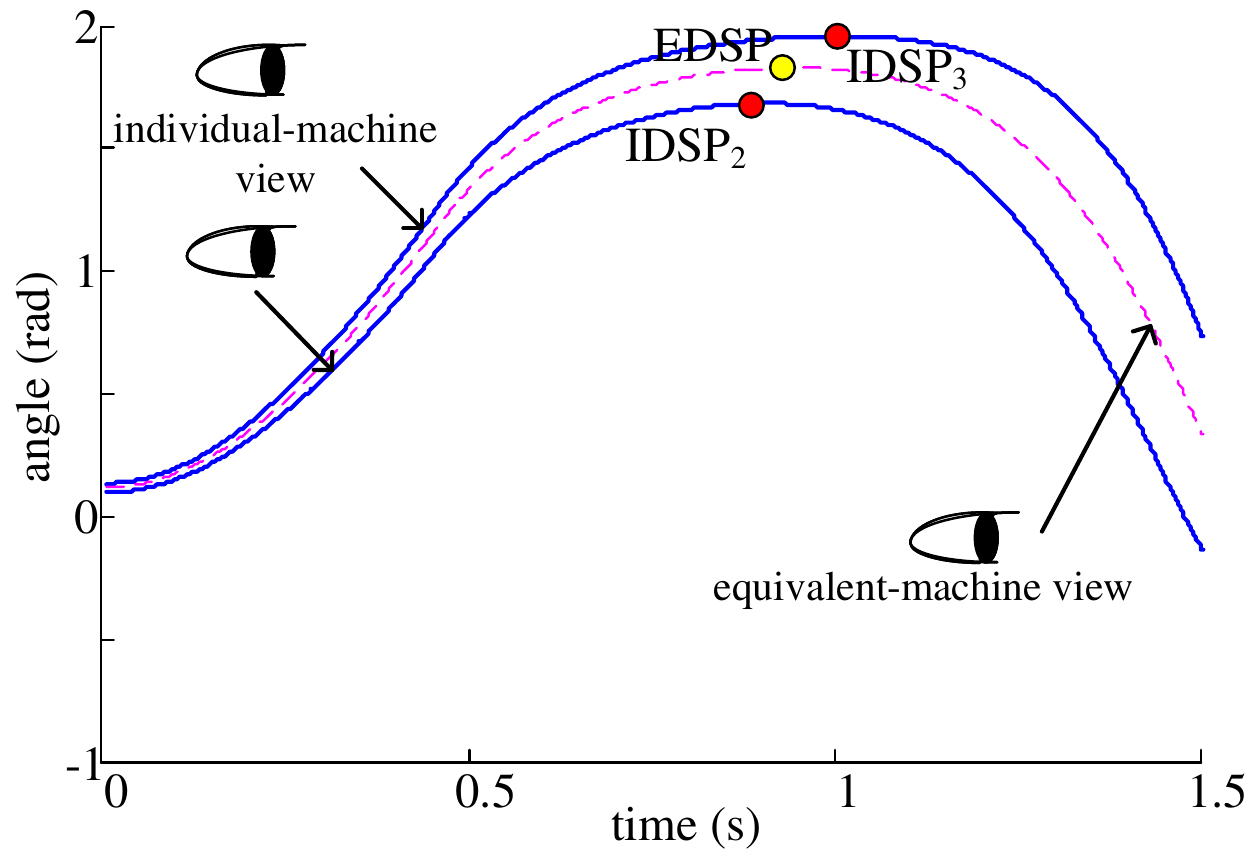}
  \caption{Occurrence of EDSPCR-SYS [TS-1, bus-4, 0.446 s].} 
  \label{fig8}  
\end{figure}  
\vspace*{-1em}
\begin{figure}[H]
  \centering
  \includegraphics[width=0.4\textwidth,center]{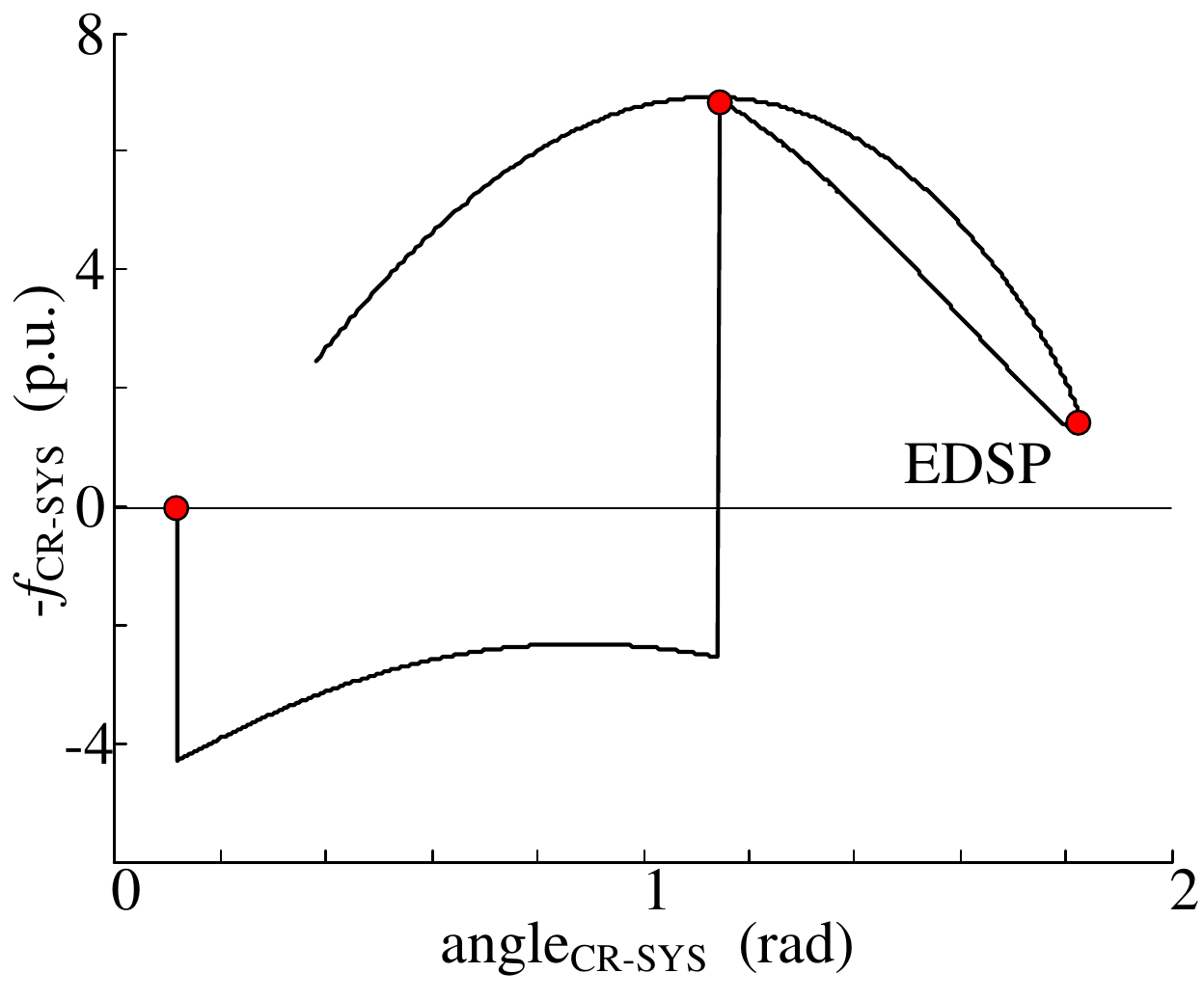}
  \caption{Kimbark curve of Machine-CR [TS-1, bus-4, 0.446 s].} 
  \label{fig9}  
\end{figure}
\vspace*{-1em}
\begin{table}[H]
  \caption{Velocity of the Equivalent machine}
  \centering
  \begin{tabular}{@{}cccc@{}}
  \toprule
  \begin{tabular}[c]{@{}c@{}}Machine \\ NO.\end{tabular} & \begin{tabular}[c]{@{}c@{}}Velocity at\\ $\text{IDSP}_2$ (p.u.)\end{tabular} & \begin{tabular}[c]{@{}c@{}}Velocity at\\ EDSP (p.u.)\end{tabular} & \begin{tabular}[c]{@{}c@{}}Velocity at\\ $\text{IDSP}_3$ (p.u.)\end{tabular} \\ \midrule
  Machine 2                                              & {\ul 0.0000}                                                       & -0.0005                                                           & -0.0014                                                            \\
  Machine-CR                                             & 0.0004                                                             & {\ul 0.0000}                                                      & -0.0007                                                            \\
  Machine 3                                              & 0.0007                                                             & 0.0004                                                            & {\ul 0.0000}                                                        \\ \bottomrule
  \end{tabular}
  \label{table2}
\end{table}
\vspace*{-0.5em}
From Fig. \ref{fig8} and Table \ref{table2}, $\text{EDSP}_{\text{CR-SYS}}$ occurs at 0.944 s. It just occurs between $\text{IDSP}_2$ and $\text{IDSP}_3$ along time horizon.
At this moment, Machine 2 already inflects back with negative velocity in its second swing, while Machine 3 is still moving forward with positive velocity in its first swing.
Therefore, $\text{EDSP}_{\text{CR-SYS}}$ can be seen as the equivalence of the IDSPs of all the stable critical machine in Group-CR.

\subsection{OCCURRENCE OF EDLP  (CHARACTERISTIC-III)} \label{section_IIID}
\noindent \textit{Explanation}: The EDLP can be seen as the equivalence of the IDLPs of all the unstable machines in Group-CR.
\\
\textit{Analysis}: Once $\text{DLP}_{\text{CR-SYS}}$ occurs along time horizon, $f_{\text{CR-SYS}}$ will reach zero, and thus the following holds
\begin{equation}
  \label{equ15}
  \sum_{i \in \Omega_{\mathrm{CR}}} f_{i\mbox{-}\mathrm{SYS}}=f_{\mathrm{CR}\mbox{-}\mathrm{SYS}}=0
\end{equation}
\par Eq. (\ref{equ15}) indicates the following
\vspace*{0.5em}
\\
(i) $f_{\text{CR-SYS}}$ can not reach zero if all $f_{i\mbox{-}\text{SYS}}$ are positive.\\
(ii) $f_{\text{CR-SYS}}$ can not reach zero if all $f_{i\mbox{-}\text{SYS}}$ are negative. \\
(iii) $f_{\text{CR-SYS}}$ may reach zero only when some $f_{i\mbox{-}\text{SYS}}$ become negative while the rest still remains positive.
\vspace*{0.5em}
\par From analysis above, $\text{EDLP}_{\text{CR-SYS}}$ will occur among the IDLPs of all machines in $\Omega_{\mathrm{CR}}$ along time horizon.
At the moment that $\text{EDLP}_{\text{CR-SYS}}$ occurs, some critical machines in $\Omega_{\mathrm{CR}}$ already go unstable with positive $f_{i\mbox{-}\text{SYS}}$, while the other critical machines in $\Omega_{\mathrm{CR}}$ are still moving forward in their first swings with negative $f_i$.
\par
\begin{figure}[H]
  \centering
  \includegraphics[width=0.4\textwidth,center]{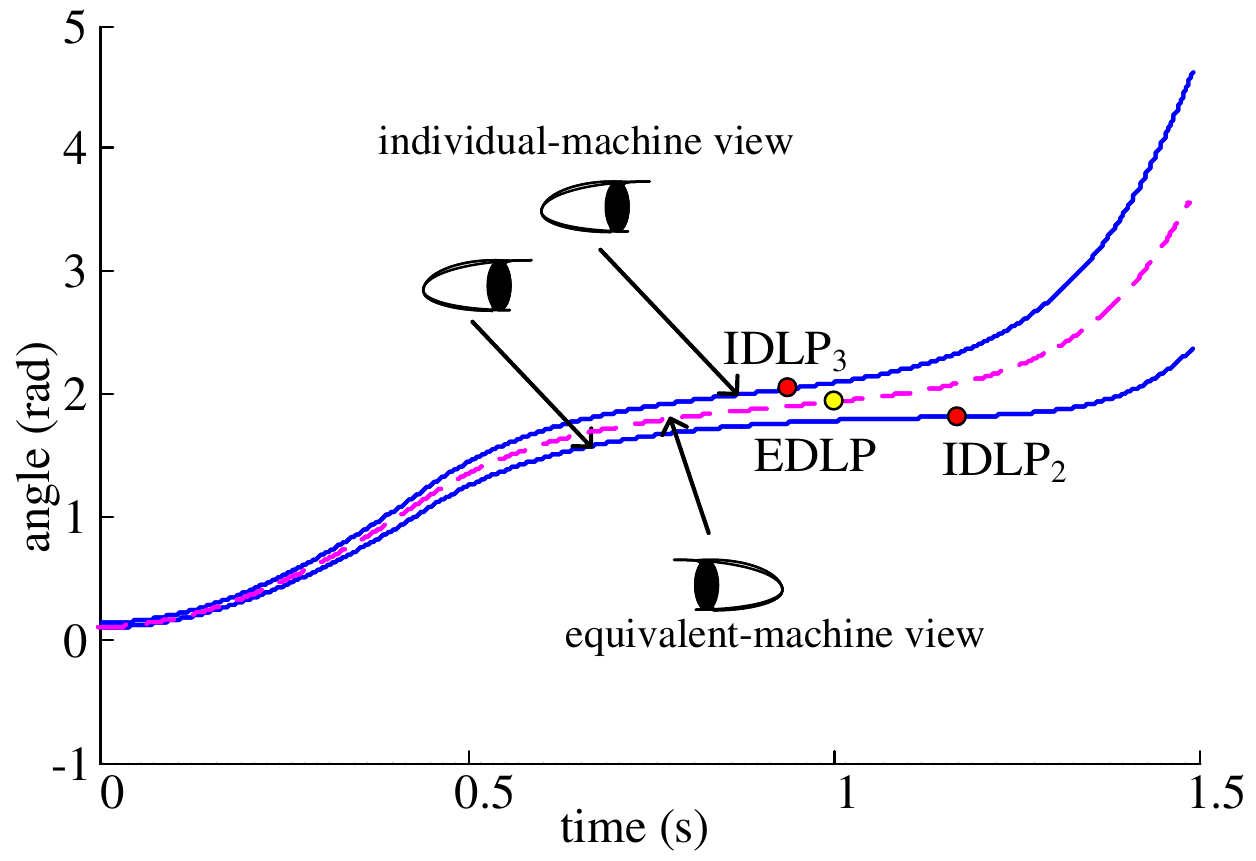}
  \caption{Occurrence of EDLPCR [TS-1, bus-4, 0.447 s].} 
  \label{fig10}  
\end{figure}

\begin{figure}[H]
  \centering
  \includegraphics[width=0.4\textwidth,center]{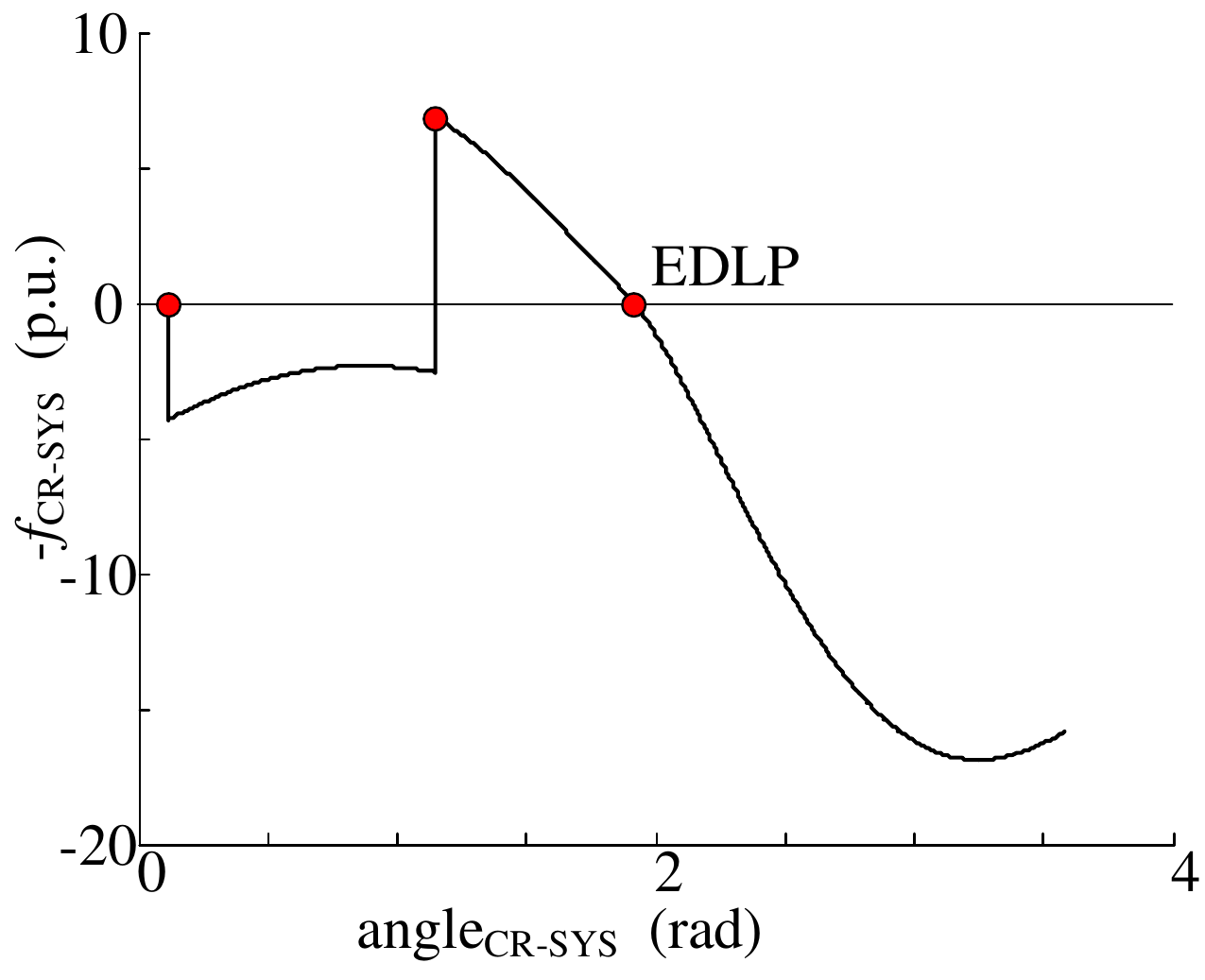}
  \caption{Kimbark curve of Machine-CR [TS-1, bus-4, 0.447 s].} 
  \label{fig11}  
\end{figure}

 The occurrence of $\text{EDLP}_{\text{CR-SYS}}$ is demonstrated in Fig. \ref{fig10}. In this case both Machines 2 and 3 become critical stable. The occurrence of the IDLP of each machine is also shown in the figure. The Kimbark curve of Machine-CR in the COI-SYS reference is shown in Fig. \ref{fig11}. 
The $f_i$ of the each critical machine at $\text{EDLP}_{\text{CR-SYS}}$ are shown in Table \ref{table3}.
\vspace*{-1em}
\vspace*{-1em}
\begin{table}[H]
  \caption{\textit{f} of the Equivalent machine}
  \centering
  \begin{tabular}{@{}cccc@{}}
  \toprule
  \begin{tabular}[c]{@{}c@{}}Machine \\ NO.\end{tabular} & \begin{tabular}[c]{@{}c@{}}\textit{f} at\\ $\text{IDLP}_3$ (p.u.)\end{tabular} & \begin{tabular}[c]{@{}c@{}}\textit{f} at\\ $\text{EDLP}_{\text{CR}}$ (p.u.)\end{tabular} & \begin{tabular}[c]{@{}c@{}}\textit{f} at\\ $\text{IDLP}_2$ (p.u.)\end{tabular} \\ \midrule
  Machine 3                                              & {\ul 0.0000}                                                       & 0.2911                                                           & 1.8811                                                            \\
  Machine-CR                                             & -0.3488                                                             & {\ul 0.0000}                                                      & 1.8811                                                            \\
  Machine 2                                              & -0.3488                                                             & -0.2911                                                            & {\ul 0.0000}                                                        \\ \bottomrule
  \end{tabular}
  \label{table3}
\end{table}
\vspace*{-0.5em}
From Fig. \ref{fig10} and Table \ref{table3}, $\text{EDLP}_{\text{CR-SYS}}$ occurs at 0.945 s. It just occurs between $\text{IDLP}_2$ and $\text{IDLP}_3$ along time horizon. At this moment, $f_3$ already becomes positive while $f_2$ is still negative.
Therefore, $\text{EDLP}_\text{CR-SYS}$ can be seen as the equivalence of the IDLPs of all the unstable critical machine in Group-CR.
\par From the analysis in this section, all the transient characteristics of the equivalent machine are essentially based on the ``motion equivalence”.
This motion equivalence ensures the establishment of the equivalent equation of motion in the equivalent machine. Against this background, the equivalent machine strictly follows all the paradigms and it will show advantages in TSA. This will be analyzed in the following section.

\section{EQUIVALENT SYSTEM STABILITY} \label{section_IV}
\subsection{FOLLOWINGS OF THE MACHINE PARADIGMS} \label{section_IVA}
Based on the analysis in Sections \ref{section_II} and \ref{section_III}, the equivalent-machine strictly follows all the paradigms. These strict followings are given as below
\\
\textit{Followings of the trajectory paradigm}: From Section \ref{section_IIA}, the system engineer monitors the EMTR of Machine-CR in the COI-NCR reference, i.e., the separation of Machine-CR with respect to the Machine-NCR. Therefore, this EMTR monitoring strictly follows the trajectory paradigm.
\\
\textit{Followings of the modeling paradigm}: From Section \ref{section_IIB}, the EMTR is modeled through the corresponding CR-NCR system that is formed by Machine-CR and Machine-NCR. Therefore, this CR-NCR system modeling strictly follows the modeling paradigm.
\\
\textit{Following of the energy paradigm}: From Section \ref{section_IIC}, the EMTE is defined in a Newtonian energy manner. Therefore, this Newtonian definition of the EMTE strictly follows the energy paradigm. EMEAC also holds in the equivalent-machine based TSA.
\par In brief, all the strict followings of the machine paradigms of the equivalent machine are fully ensured by the ``motion equivalence” of all the individual machines inside each group, as analyzed in Section \ref{section_III}. This is because the equation of motion is established in the equivalent machine through this motion equivalence. Note that all these strict followings only hold in the equivalent system.

\subsection{EQUIVALENT MACHINE STABILITY AND THE SYSTEM STABILITY} \label{section_IVB}
\noindent\textit{Equivalent machine stability under a possible pattern}: Following the analysis in Section \ref{section_IIC}, for a given group separation pattern, the stability margin of the equivalent machine is denoted as
\begin{equation}
  \label{equ16}
  \begin{split}
    \eta_{\mathrm{CR}\mbox{-}\mathrm{SYS}}&=\frac{\Delta V_{P E \mathrm{C R}\mbox{-}\mathrm{S Y S}}-V_{K E \mathrm{C R}\mbox{-}\mathrm{S Y S}}^{c}}{V_{K E \mathrm{C R}\mbox{-}\mathrm{S Y S}}^{c}}\\
    &=\frac{A_{\mathrm{decCR}\mbox{-}\mathrm{SYS}}-A_{\mathrm{accCR}\mbox{-}\mathrm{SYS}}}{A_{\mathrm{accCR}\mbox{-}\mathrm{SYS}}}
  \end{split}
\end{equation}
\par
\begin{figure}[H]
  \centering
  \includegraphics[width=0.47\textwidth,center]{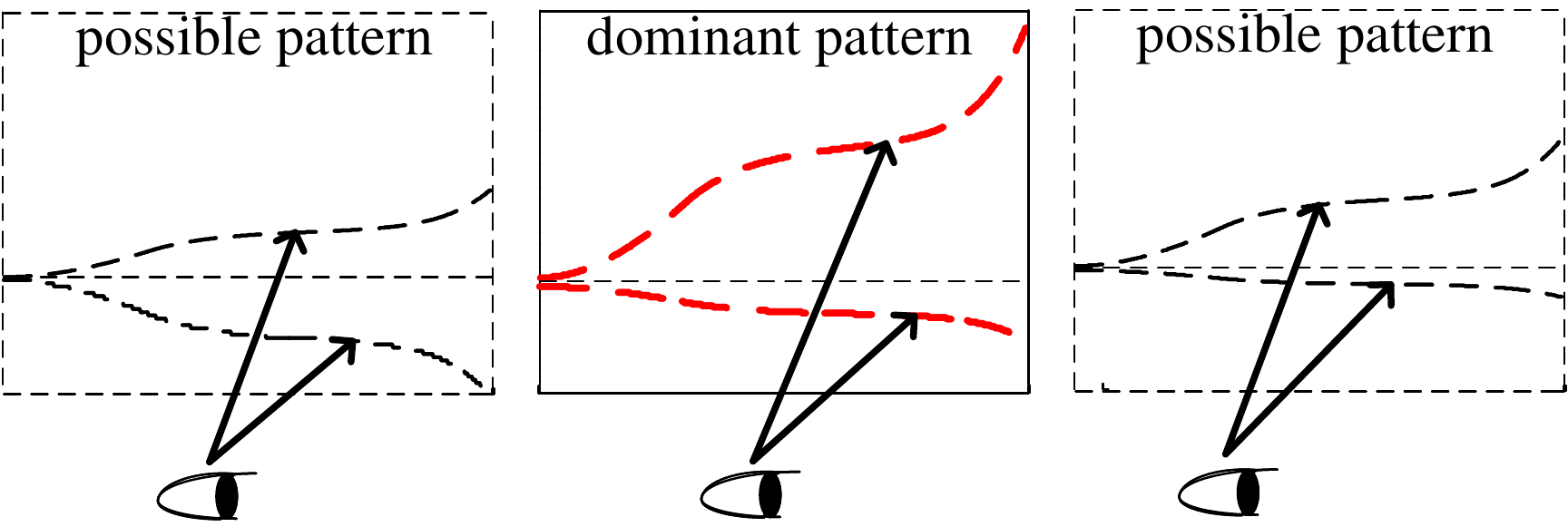}
  \caption{Group-separation-pattern monitoring.} 
  \label{fig12}  
\end{figure}

 From Eq. (\ref{equ16}), the stability state of the equivalent machine can be characterized through the sign of $\eta_{\mathrm{CR}\mbox{-}\mathrm{NCR}}$:
$\eta_{\mathrm{CR}\mbox{-}\mathrm{NCR}}>0$ means that the machine is stable; $\eta_{\mathrm{CR}\mbox{-}\mathrm{NCR}}=0$ means that the machine is critical stable; and $\eta_{\mathrm{CR}\mbox{-}\mathrm{NCR}}<0$ means that the machine becomes unstable.
Note that the stability margin, i.e., the ``severity” of the machine is measured through the absolute value of $\eta_{\mathrm{CR}\mbox{-}\mathrm{NCR}}$.
\\ \textit{Dominant group separation pattern}: Following the analysis in Section \ref{section_IIA}, for a multi-machine system, the system operators need to monitor all the possible group separation patterns. This group-separation-pattern monitoring is shown in Fig. \ref{fig12}.
\vspace*{-0.5em}
From Fig. \ref{fig12}, among all the possible group separation patterns, the dominant separation pattern is defined as the pattern with the lowest margin among all the possible patterns
\begin{equation}
  \label{equ17}
  \eta_{\mathrm{CR}\mbox{-}\mathrm{SYS}}^{(\mathrm{dom})}=\min \left\{\eta_{\mathrm{CR}\mbox{-}\mathrm{SYS}}^{(i)}\right\}
\end{equation}
\par From Eq. (\ref{equ17}), the dominant pattern reflects the most severe separation among all the possible patterns. Based on this, the motion of the Machine-CR under the dominant pattern is seen as the ``one-and-only” crucial motion in TSA among all the possible cases. Note that all the equivalent-machine analysis in the following paper is given under the dominant pattern for simplicity and clearance.
\\
\textit{Equivalent system stability}: Following Eq. (\ref{equ17}), the equivalent system under the dominant pattern is certain to be the closest one to the original system. Therefore, the stability margin of the equivalent system is given as
\begin{equation}
  \label{equ18}
  \eta_{\mathrm {sys }}=\eta_{\mathrm{CR}\mbox{-}\mathrm{SYS}}^{(\mathrm {dom })}
\end{equation}
\par Eq. (\ref{equ18}) indicates that the stability and severity of the equivalent system will be obtained simultaneously because the equivalent machine is the ``one-and-only” machine in TSA.

\subsection{FURTHER ANALYSIS OF THE EQUIVALENT MACHINE} \label{section_IVC}
From the analysis in Section \ref{section_II}, the equivalent-machine analyst focuses on the transient characteristic of the one-and-only equivalent machine in the equivalent system.
Based on this equivalent-machine monitoring, the equivalent equation of motion inside the equivalent machine is established. Against this background, the one-and-only CR-NCR system with strict NEC characteristic is established. The equivalent machine can be seen as the special ``equivalent” case of the individual-machine.
Compared with the superimposed-machine that is modeled based on the ``energy superimposition” \cite{3}, the equivalent machine is based on the ``motion equivalence”. This motion equivalence fully ensures the strict followings of the machine paradigms in the equivalent-machine.
\par
\begin{figure}[H]
  \centering
  \includegraphics[width=0.45\textwidth,center]{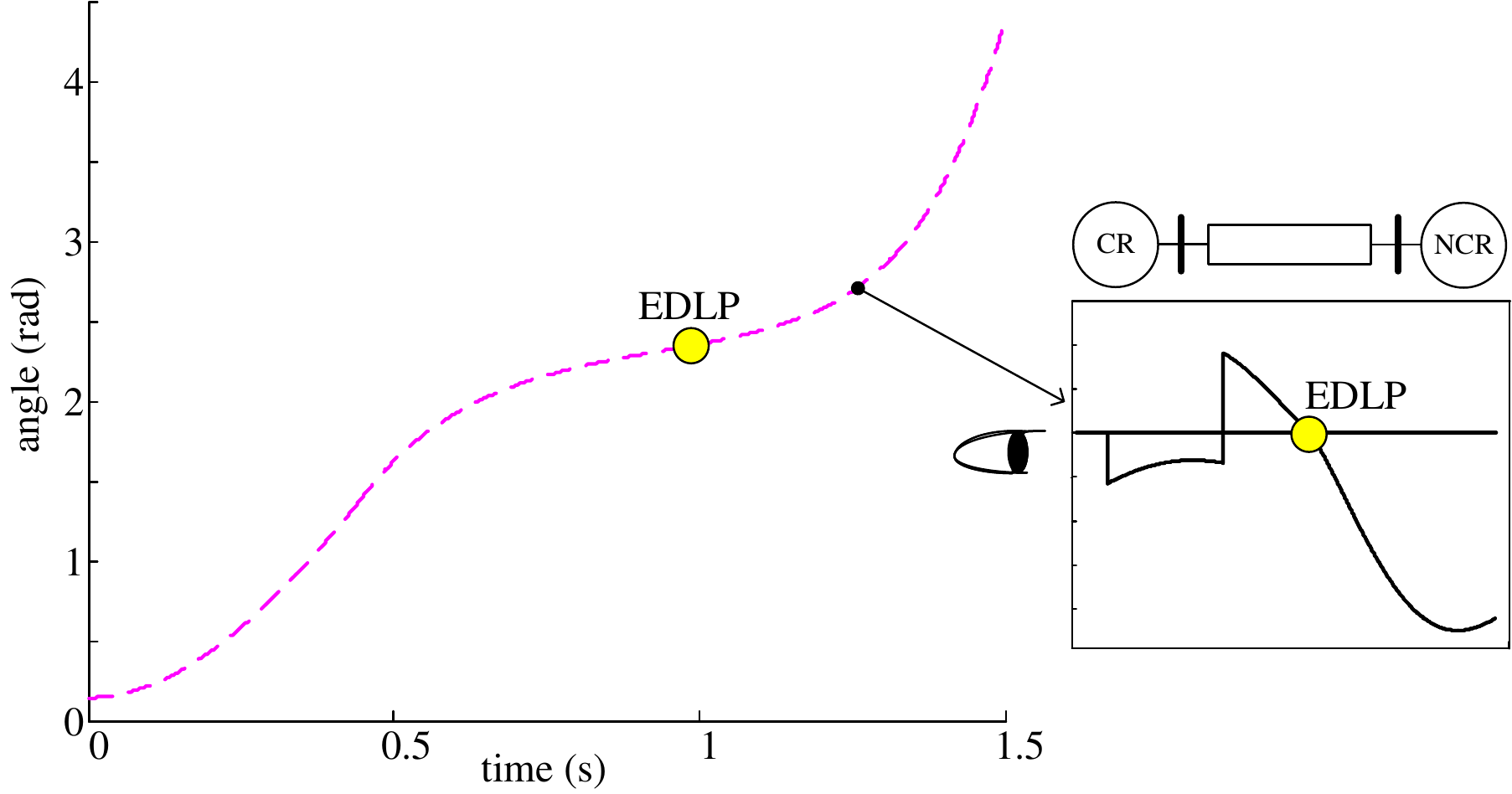}
  \caption{The use of the equivalent machine in the equivalent system [TS-1, bus-4, 0.447 s].} 
  \label{fig13}  
\end{figure}

 The use of the equivalent machine in the equivalent system is shown in Fig. \ref{fig13}.
\vspace*{-0.5em}
The strict followings of the paradigms essentially ensure the advantages of the equivalent machine and also its application in TSA. The advantages are given as below:
\\
\noindent Stability-characterization advantage: The stability of EMTR of the one-and-only equivalent machine is characterized precisely at EMPP.
\\
Trajectory-depiction advantage: The variance of EMTR of the one-and-only equivalent machine is depicted clearly through the individual-machine EMPP.
\vspace*{0.5em}
\par The two advantages will be fully reflected in the definitions of the equivalent-machine based transient stability concepts. The authors emphasize again that these advantages only hold in the equivalent system. Detailed analysis is given in Section \ref{section_V}.

\section{PRECISE DEFINITIONS OF THE EQUIVALENT-MACHINE BASED TRANSIENT STABILITY CONCEPTS} \label{section_V}
\subsection{EQUIVALENT MACHINE SWING} \label{section_VA}
\noindent\textit{Statement}: The trajectory-depiction advantage is fully reflected in the definition of the equivalent-machine swing.
\\
\textit{Equivalent machine perspective}: The system engineer monitors the ``one-and-only” EMTR (under dominant group separation pattern) in the equivalent system. The definitions of the stable and unstable critical-machine swing are given as below
\\
Swing of a stable equivalent machine: It is defined as the EDSP of the machine. This is because EDSP reflects the ``inflection” of the EMTR of the machine ($\mathrm{d}\delta_{\mathrm{CR}\mbox{-}\mathrm{SYS}}/\mathrm{d}t=\omega_{\mathrm{EDSP}}=0$), as shown in Fig. \ref{fig10}.
\\
Swing of an unstable equivalent machine: It is defined as the EDLP of the machine. This is because EDLP reflects the ``separation” of the IMTR of the machine ($\mathrm{d}^2\delta_{\mathrm{CR}\mbox{-}\mathrm{SYS}}/\mathrm{d}t^2=f_{\mathrm{EDLP}}=0$), as shown in Fig. \ref{fig12}.
\par Based on the definitions above, the concept of the equivalent machine swing focuses on the depiction of the trajectory variance of the ``one-and-only” equivalent machine in the equivalent system. Therefore, this equivalent-machine swing is also seen as the ``system swing” of the equivalent system.
\\
\textit{Example}: Tutorial examples are already given in Figs. \ref{fig8} and \ref{fig10}. From the two figures, the EMTR variance of equivalent machine is depicted clearly at EMPP.

\subsection{CRITICAL STABILITY OF THE EQUIVALENT SYSTEM} \label{section_VB}
\begin{figure}[H]
  \centering
  \includegraphics[width=0.37\textwidth,center]{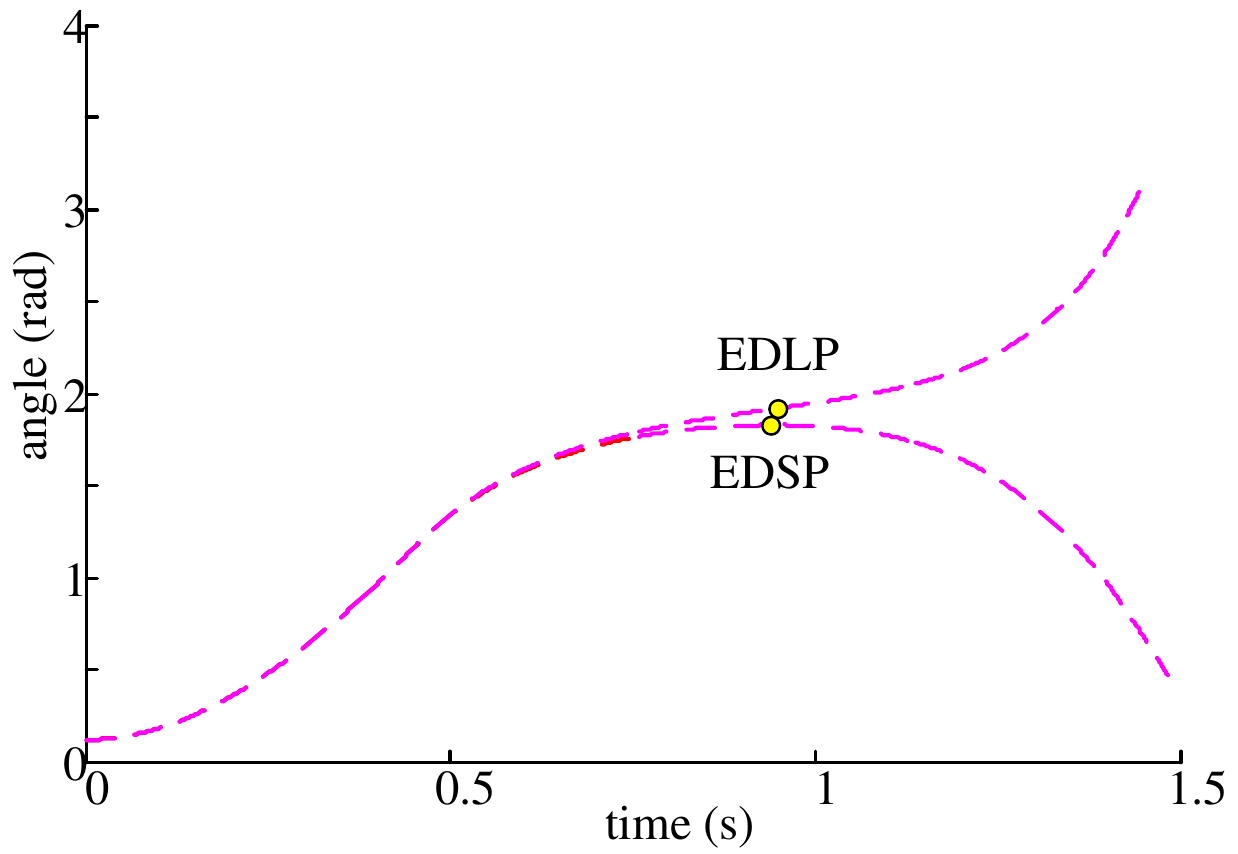}
  \caption{Critical stable and critical unstable equivalent system trajectories [TS-1, bus-4, 0.447s].} 
  \label{fig14}  
\end{figure}

\noindent\textit{Statement}: Both the stability-characterization advantage and the trajectory-depiction advantage are reflected in the definition of the critical stability o f the equivalent system.
\\
\textit{Equivalent machine perspective}: The critical stability state of the equivalent system is completely decided through the critical stability of the ``one-and-only” equivalent machine. Therefore, the IDSP of the critical stable equivalent machine becomes of key value in the critical stability analysis of the equivalent system. 
\\
\textit{Example}: The simulation cases of the critical stable and critical unstable equivalent system trajectories are already shown in Figs. \ref{fig10} and \ref{fig12}, respectively. The Kimbark curves of the equivalent machine in the two cases are already shown in Figs. \ref{fig11} and \ref{fig13}, respectively. They two equivalent-machine trajectories are also shown in Fig. \ref{fig14} for clear comparison.
\vspace*{-0.5em}
The analysis of the critical stability of the equivalent system is given as below
\vspace*{0.5em}
\\
Equivalent machine monitoring: Following trajectory stability theory, through the comparison between the critical stable case and the critical unstable case as given in Fig. \ref{fig14}, the critical trajectory stability of the equivalent system is completely decided by the critical trajectory stability of Machine-CR.
\\
CR-SYS system modeling: According to the modeling paradigm, the critical trajectory stability of Machine-CR is modeled through the corresponding CR-SYS system.
\\
EMEAC: According to the energy paradigm, the one-and-only Machine-CR in the equivalent system maintains critical stable when $t_c$ is 0.446 s, and it becomes critical unstable when $t_c$ is 0.447 s.
\vspace*{0.5em}
\par From analysis above, the equivalent machine shows two advantages in the definitions of the critical stability of the equivalent system.
\\
(i) The stability state of the equivalent machine from the critical stability to the critical instability is characterized precisely through the change from $\text{EDSP}_{\text{CR}}$ to the $\text{EDLP}_{\text{CR}}$, as in Figs. \ref{fig11} and \ref{fig13}.
\\(ii) The trajectory variance of the equivalent machine from the critical stability to the critical instability is depicted clearly through the change from $\text{EDSP}_{\text{CR}}$ to the $\text{EDLP}_{\text{CR}}$, as in Fig. \ref{fig14}.
\par (i) and (ii) are fully based on the strict followings of the machine paradigms in the equivalent machine. Note that the two advantages only hold in the equivalent system.

\subsection{DISAPPEARANCE OF THE EQUIVALENT-MACHINE PES} \label{section_VC}
The concept of the equivalent-machine potential energy surface does \textit{not} exist. This is because the EMTR of the equivalent machine is depicted in the ``two-dimensional” time-angle space. Against this background, the potential energy surface that requires at least three dimensions is unable to be modeled \cite{2}, \cite{3}.

\section{CASE STUDY} \label{section_VI}
\begin{figure}[H]
  \centering
  \includegraphics[width=0.42\textwidth,center]{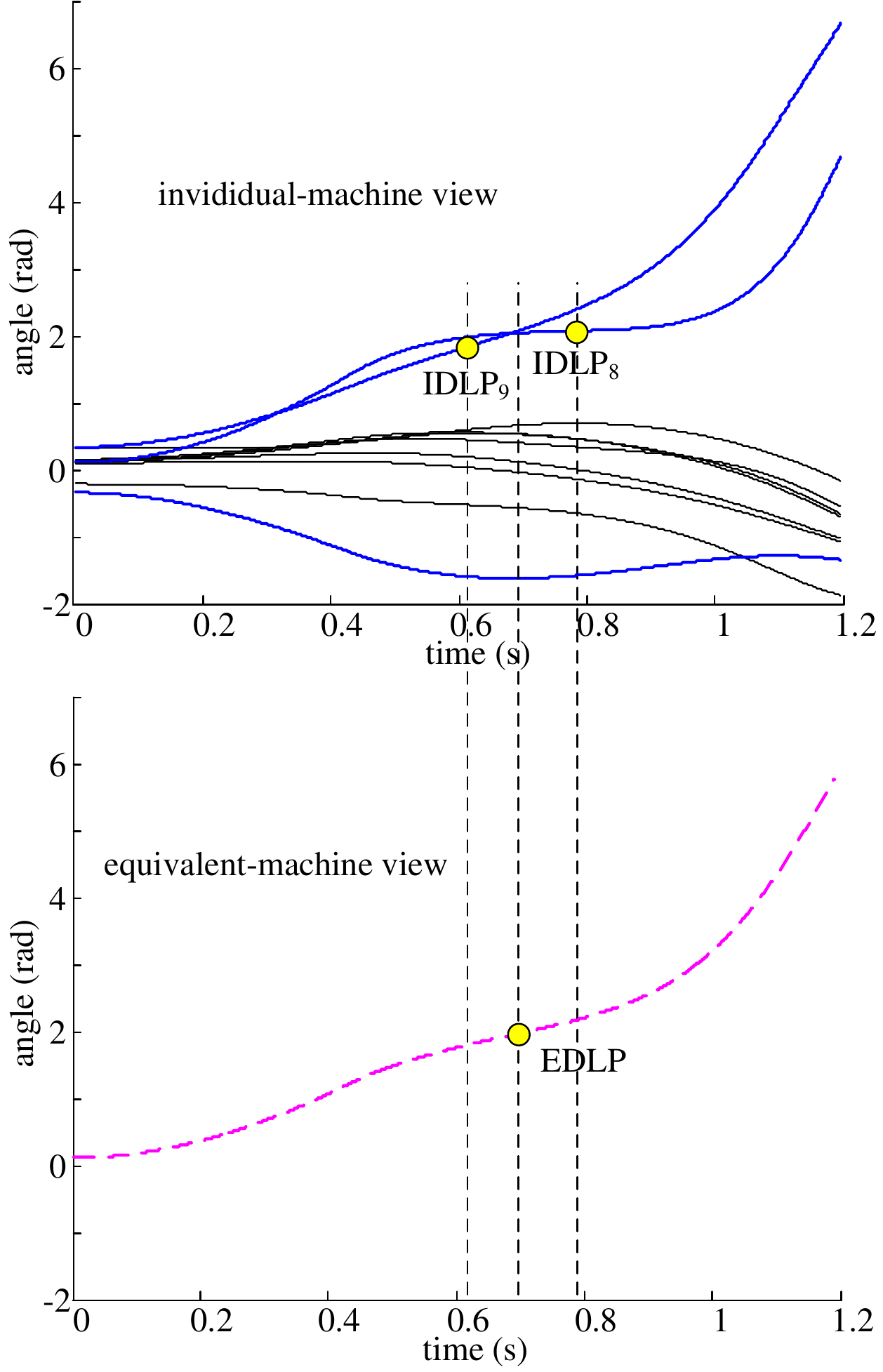}
  \caption{Comparison between individual-machine and equivalent-machine [TS-1, bus-2, 0.430 s].} 
  \label{fig15}  
\end{figure}

\begin{figure}[H]
  \centering
  \includegraphics[width=0.4\textwidth,center]{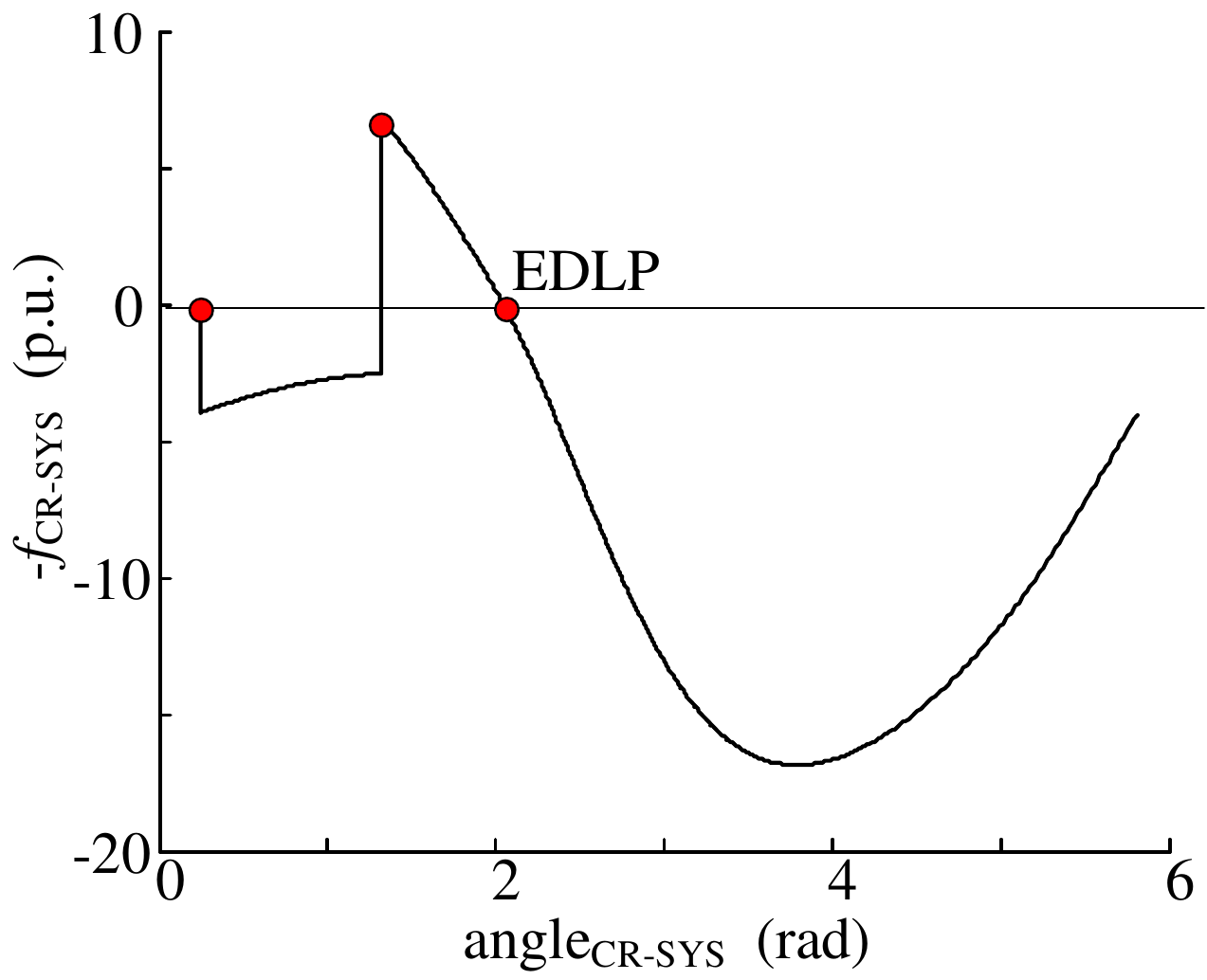}
  \caption{Kimbark curve of Machine-CR [TS-1, bus-2, 0.430 s].} 
  \label{fig16}  
\end{figure}

The case [TS-1, bus-2, 0.430 s] is provided here to show the comparisons between the equivalent-machine method and the individual-machine in TSA.
In this case $\Omega_{\mathrm{CR}}$ is \{Machine 8, Machine 9\}. Both the individual-machine based TSA and the equivalent-machine based TSA are shown in Fig. \ref{fig15}.
The Kimbark curves of Machines 8 and 9 were already given in Ref. \cite{1}. The Kimbark curve of Machine-CR (in the COI-SYS reference) is shown in Fig. \ref{fig16}. Note that the mirror system is used in this simulation case.
\vspace*{-1em}
\vspace*{-0.5em}
\noindent 1) INDIVIDUAL-MACHINE BASED TSA
\vspace*{0.5em}
\par After fault clearing, the individual-machine analyst only monitors Machines 8, 9 and 1 because they are critical machines. Along the post-fault original system trajectory, the individual-machine analyst may focuses on the following instants.
\vspace*{0.5em}
\\
\textit{$IDLP_9$ occurs (0.614 s)}: Machine 9 is judged as unstable, and $\eta_9$ is computed as -0.594.
\\
\textit{$IDSP_{39}$ occurs (0.686 s)}: Machine 39 is judged as stable.
\\
\textit{$IDLP_8$ occurs (0.776 s)}: Machine 8 is judged as unstable, and $\eta_8$ is computed as -0.002.
\vspace*{0.5em}
\par The stability evaluation of the original system is given as below.
\vspace*{0.5em}
\\
\textit{$IDLP_9$ occurs (0.614 s)}: The stability of the system is characterized as becoming unstable according to the unity principle.
\\
\textit{$IDLP_8$ occurs (0.776 s)}: The severity of the system is obtained. $\eta_{\mathrm{sys}}$=[-0.594, -0.002].
\vspace*{0.5em}
\par From analysis above, based on the critical-machine monitoring of the original system, the stability of the original system is evaluated in a ``machine-by-machine” manner. The stability and severity of the system is obtained at $\text{DLP}_9$ and $\text{DLP}_8$, respectively.
\vspace*{0.5em}
\\
2) EQUIVALENT-MACHINE BASED TSA
\vspace*{0.5em}
\par After fault clearing, the equivalent-machine analyst only monitors the ``one-and-only” Machine-CR in the equivalent machine in the system. Along the post-fault equivalent system trajectory, the equivalent-machine analyst focuses on the following instants.
\vspace*{0.5em}
\\
\textit{$EDLP_{CR}$ occurs (0.686 s)}: Machine-CR is judged as unstable, and $\eta_{\mathrm{CR}}$ is computed as -0.198. At the moment, the equivalent system is also evaluated to become unstable. $\eta_{\mathrm{sys}}=\eta_{\mathrm{CR}}$.
\vspace*{0.5em}
\par From analysis above, based on the ``one-and-only” equivalent-machine monitoring of the equivalent system, the stability of the equivalent system is evaluated in a ``one-and-only machine” manner. The stability and severity of the system is obtained at $\text{EDLP}_{\text{CR}}$ simultaneously.
\vspace*{0.5em}
\\
3) COMPARISON BETWEEN INDIVIDUAL-MACHINE AND EQUIVALENT MACHINE
\vspace*{0.5em}
\par
\begin{figure}[H]
  \centering
  \includegraphics[width=0.4\textwidth,center]{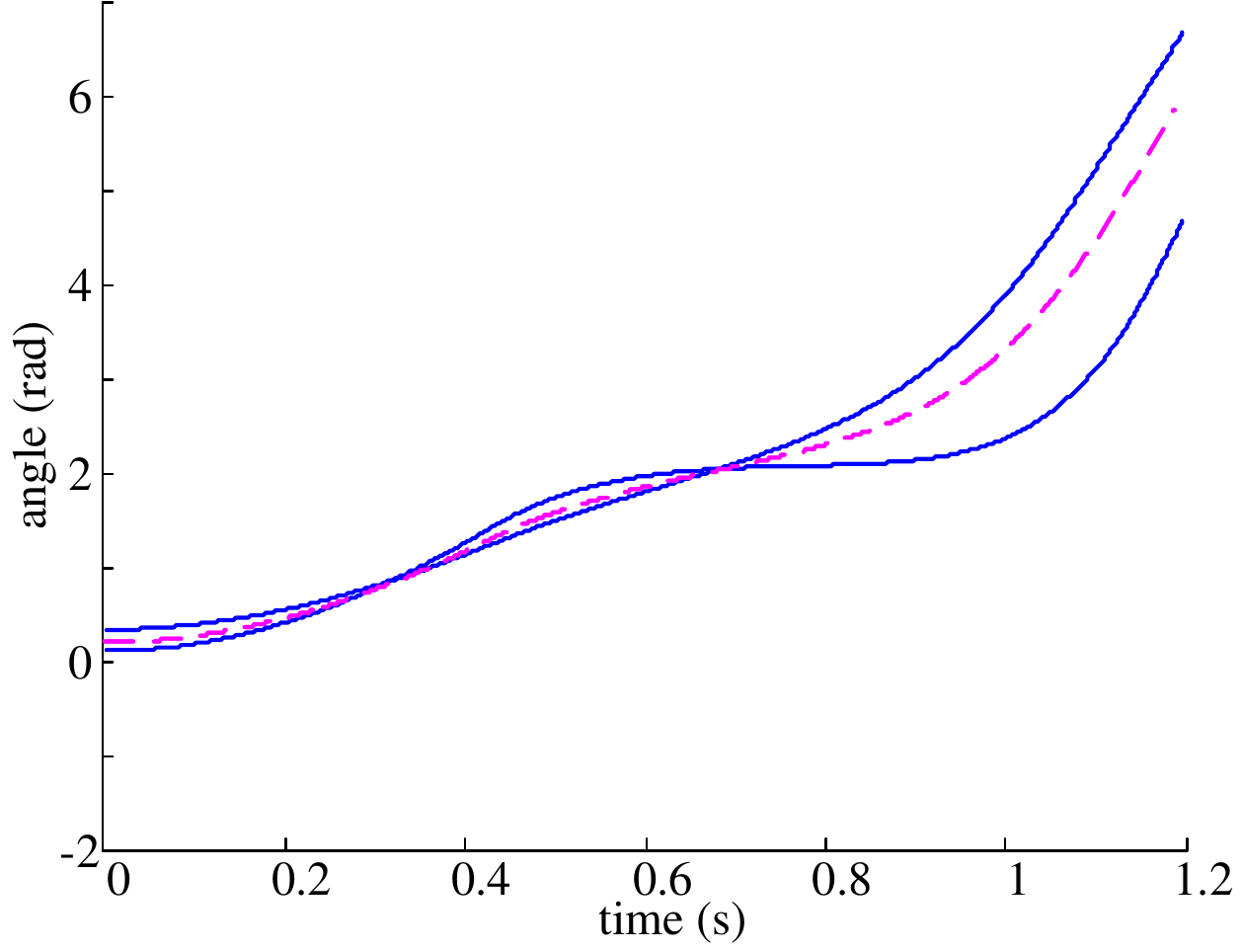}
  \caption{Difference between the original system trajectory and the equivalent system trajectory [TS-1, bus-2, 0.430 s].} 
  \label{fig17}  
\end{figure}

 Comparison between the original system trajectory and the equivalent system trajectory is shown in Fig. \ref{fig17}.
\vspace*{-0.5em}
From analysis above, compared with the individual-machine case, the equivalent-system stability evaluation show the following characteristics.
\vspace*{0.5em}
\\
(i) The ``stability” of the equivalent system is characterized later than that of the original system ($\text{IDLP}_9$ occurs earlier than $\text{EDLP}_{\text{CR}}$).
\\
(ii) The “severity” of the equivalent system is characterized earlier than that of the original system ($\text{IDLP}_8$ occurs later than $\text{EDLP}_{\text{CR}}$).
\vspace*{0.5em}
\par (i) and (ii) can be seen as the reflection of the equivalence of the EDLP.
\par At this stage, revisiting the individual-machine and equivalent-machine, one can find that the two machines strictly follow the machine paradigms.
These strict followings indicate that the strict correlation between the ``trajectory variance” and ``energy conversion” can be established in the two machines, and thus the two machines will show both the stability-characterization advantage and the trajectory-depiction advantage in TSA.
However, it should be emphasized that the advantages of the individual-machine and the equivalent machine are shown in the original system and the equivalent system, respectively. In other words, the advantages of the equivalent machine can only be shown in the equivalent system, while they are unable to be found in the original system. Based on this, one question naturally emerges: 
\vspace*{0.5em}
\par \textit{Are the stabilities and severities of the two systems identical}?
\vspace*{0.5em}
\par This question is quite complicated. This is because the stabilities of the two systems are completely identical, while the severities of the two systems might be different under certain circumstances. Clarifications are given in the following section.

\section{DIFFERENCES BETWEEN ORIGINAL SYSTEM AND EQUIVALENT SYSTEM} \label{section_VII}
\subsection{INNER-GROUP MACHINE MOTION}  \label{section_VIIA}
At the beginning of this section, the authors emphasize the primary and also crucial concept in the power system transient stability
\vspace*{0.5em}
\par \textit{The fundamental concept of the power system transient stability is defined through the ``original system”}.
\vspace*{0.5em}
\par In other words, the original system with multiple machines is the physical system that the power system transient stability is originally defined.
\par Revisiting the equivalent-machine thinking, in the equivalent-machine based TSA, the one-and-only equivalent Machine-CR is seen as the full representation of the original system. However, due to the motion equivalence, an unavoidable fact in the equivalent system can be found.
\vspace*{0.5em}
\par \textit{The equivalent system trajectory is not identical to the original system trajectory}.
\vspace*{0.5em}
\par
\begin{figure}[H]
  \centering
  \includegraphics[width=0.46\textwidth,center]{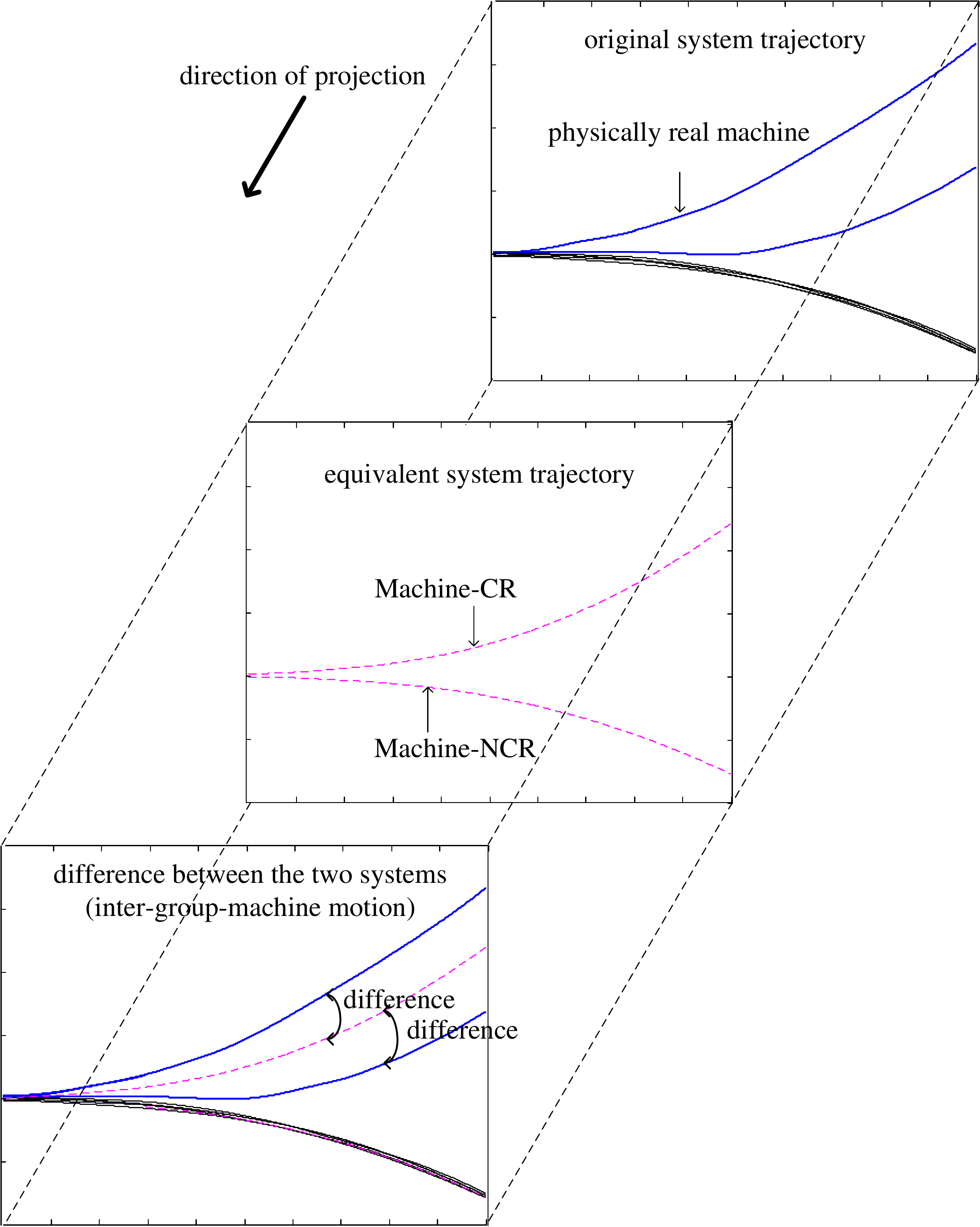}
  \caption{Demonstration of the inner-group machine motion.} 
  \label{fig18}  
\end{figure}

 This fact indicates that ``differences” always exist between the equivalent system trajectory and the original system trajectory. The differences between the original system and the equivalent system are shown in Fig. \ref{fig18}.
\vspace*{-0.5em}
\par In this paper, this difference is named the ``inner-group-machine motion”. The inner-group-machine motion is mathematically depicted as
\begin{equation}
  \label{equ19}
  \left\{\begin{array}{l}
    \delta_{i\mbox{-}\mathrm{CR}}=\delta_{i}-\delta_{\mathrm{CR}}=\delta_{i\mbox{-}\mathrm{SYS}}-\delta_{\mathrm{CR}\mbox{-}\mathrm{SYS}} \quad i \in \Omega_{\mathrm{CR}} \\
    \\
    \delta_{j\mbox{-}\mathrm{NCR}}=\delta_{j}-\delta_{\mathrm{CR}}=\delta_{j-\mathrm{SYS}}-\delta_{\mathrm{NCR}\mbox{-}\mathrm{SYS}} \quad j \in \Omega_{\mathrm{NCR}}
    \end{array}\right.
\end{equation}
\par In Eq. (\ref{equ19}), the inner-group-machine motion reflects the separation of the physically real machine with respect to the equivalent machine inside each group. Essentially, the inner-group-machine motion is ``created'' from the difference between the original system and the equivalent system. In other words, it neither exists in the original system nor be found in the equivalent system.
\par The inner-group machine motion also emerges a key challenge when using the equivalent machine method in TSA. That is
\vspace*{0.5em}
\par \textit{Could the original system be replaced with the equivalent system in TSA}?
\vspace*{0.5em}
\par In this paper, the following deductions are given
\\
(i) The stability of the equivalent system is ``identical” to that of the original system, and this is independent of the inner-group motions (clarification is given in Section \ref{section_VIIB}).
\\
(ii) The severity of the equivalent system is ``close” to that of the original system if all the inner-group machine motions are slight (clarification is given in Section \ref{section_VIIC}).
\\
(iii) The severity of the equivalent system will show significant difference with the original system if any inner-group machine motion becomes fierce (clarification is given in Section \ref{section_VIIC}).
\par (i) indicates that the system engineer may only focus on the comparison about the severities between the two systems. Further, (ii) and (iii) indicate the two most important deductions in the equivalent-machine based TSA
\vspace*{0.5em}
\par \textit{(i) The original system can be replaced with the equivalent system if all the inner-group motions are slight}.
\par \textit{(ii) This replacement will fail if any inner-group motion becomes fierce}.
\vspace*{0.5em}
\par In particular, the equivalent machine is flexible when all inner-group motions are slight, while it shows problems when any inner-group motion becomes fierce.
\par Two cases about the inner-group motions are shown in Figs. \ref{fig19} (a) and (b), respectively.
\begin{figure} [H]
  \centering 
  \subfigure[]{%
  \label{fig19a}
    \includegraphics[width=0.4\textwidth]{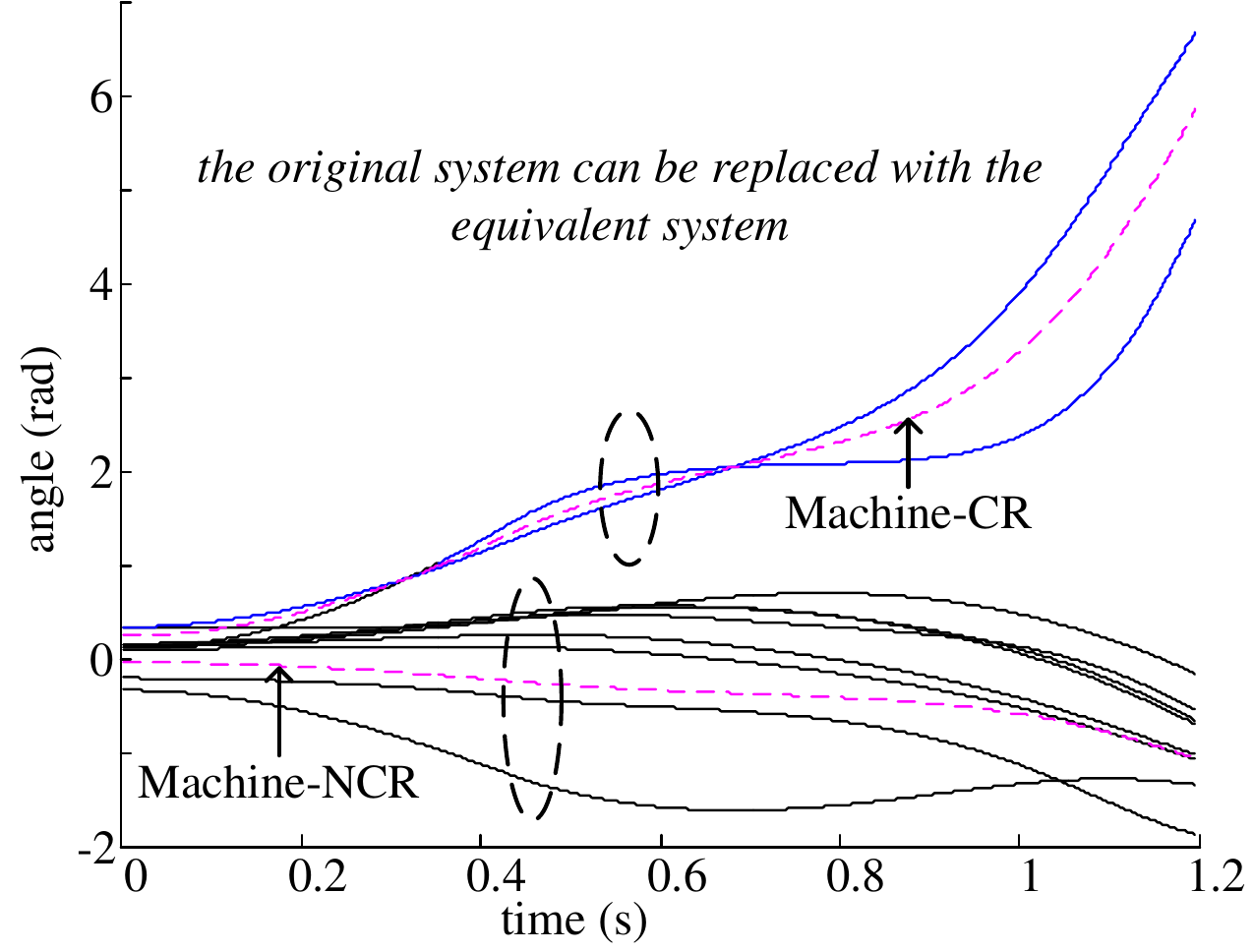}}%
\end{figure} 
\vspace*{-2em}
\addtocounter{figure}{-1}       
\begin{figure} [H]
  \addtocounter{figure}{1}      
  \centering 
  \subfigure[]{%
    \label{fig19b}
    \includegraphics[width=0.4\textwidth]{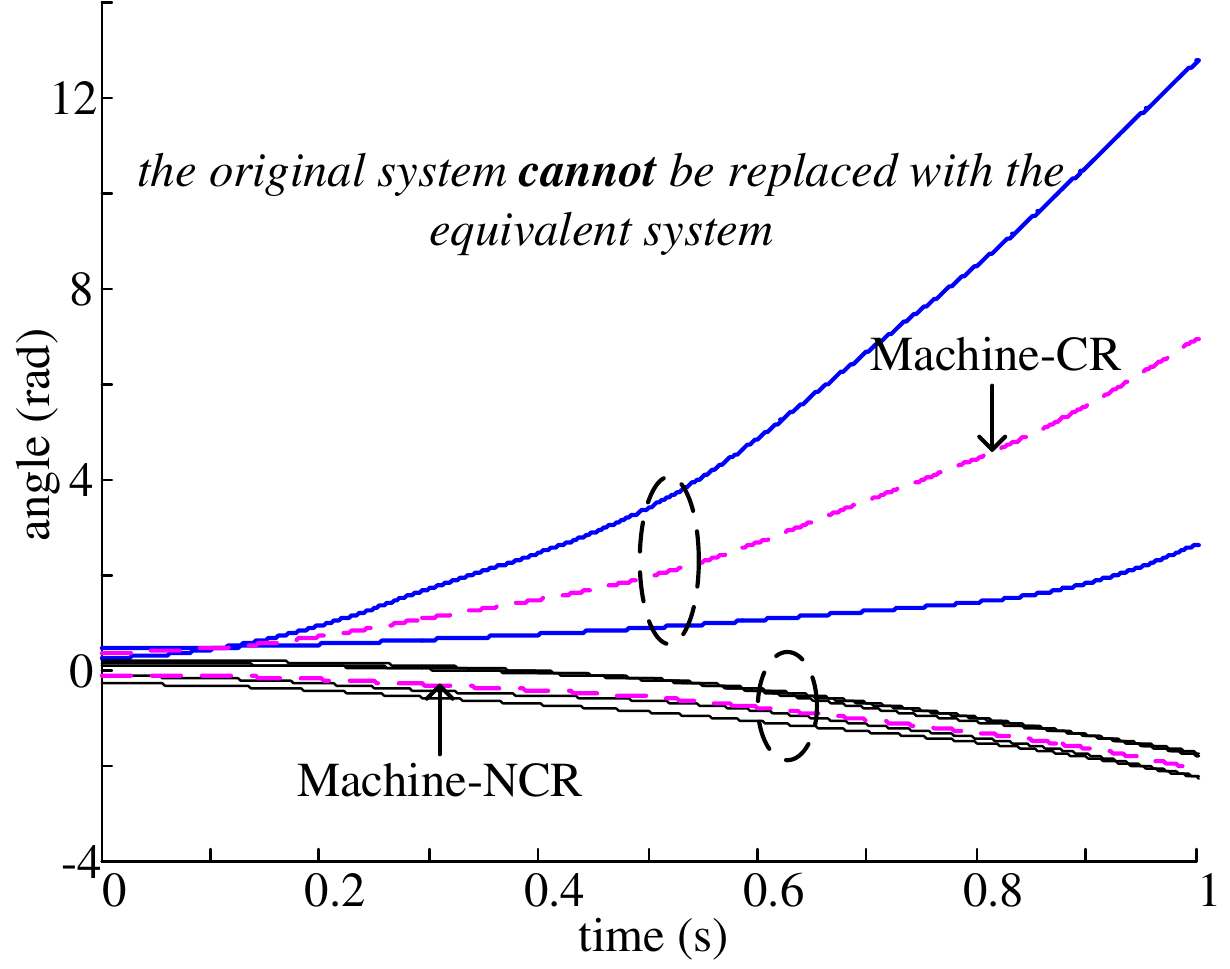}}%
  \caption{Two cases about inner-group motions. (a) Case-1: All the inner-group motions are slight  [TS-1, bus-2, 0.430 s]. (b) Case-2: Inner-group motions inside Group-CR are fierce [TS-6, bus-19, 0.260 s].}%
  \label{fig19}
\end{figure}
\vspace*{-0.5em}
In the following sections, all the deductions above will be clarified through the revisit of the definitions of the stability and severity of the original system. 

\subsection{COMPARISON OF THE SYSTEM STABILITY} \label{section_VIIB}
\noindent \textit{Definition of the original system stability}: Following the trajectory stability theory \cite{1}, the stability of the original system is depicted through the separations among the IMTRs of the machine in the COI-SYS reference. In other words, any separation of the machine in the COI-SYS reference may cause the instability of the system.
\par Based on the definition of the original system stability, the following deduction can be obtained
\vspace*{0.5em}
\par \textit{The stability of the equivalent system is identical to that of the original system, no matter how inner-group-machine motions acts}.
\vspace*{0.5em}
\\ \textit{Clarification}: In the original system, assume the IMTR of Machine \textit{i} goes infinite along time horizon. The mathematical depiction is given as
\begin{equation}
  \label{equ20}
  \left|\delta_{i\mbox{-}\mathrm{SYS}, t}\right|=\left|\int_{t_{0}}^{t} \omega_{i\mbox{-}\mathrm{SYS}} d t\right|=+\infty \quad t=+\infty
\end{equation}
\par Assume Group-CR is formed by the only one Machine \textit{i}.  Note that this is a possible pattern rather than the dominant pattern. Following Eq. (\ref{equ20}), the EMTR of Machine-CR can be given as:
\begin{equation}
  \label{equ21}
  \left|\delta_{\mathrm{CR}\mbox{-}\mathrm{SYS}, t}\right|=\left|\delta_{i\mbox{-}\mathrm{SYS}, t}\right|=+\infty \quad t=+\infty
\end{equation}
\par From the analysis above, one can obtain the following:
\\ (i) Following Eq. (\ref{equ20}), from the individual-machine perspective, the original system is evaluated to become unstable because Machine \textit{i} becomes unstable, according to the unity principle.
\\ (ii) Following Eq. (\ref{equ21}), from the equivalent-machine perspective, under the ``possible” group separation pattern, the equivalent system is evaluated to become unstable because the one-and-only Machine-CR becomes unstable.
\\ (iii) Following the definition of the ``dominant” group separation pattern as given in Eq. (\ref{equ17}), from the equivalent-machine perspective, following, the Machine-CR under the dominant pattern is certain to become unstable because it is the one with the lowest margin among all possible patterns.
\par (i) to (iii) indicates that the original system becoming unstable is identical to the equivalent system becoming unstable.
\par We further extend the analysis above into the stable case. In the original system, assume the IMTR of each machine in the system is bounded along time horizon. The mathematical depiction is given as 
\begin{equation}
  \label{equ22}
  \left|\delta_{i\mbox{-}\mathrm{SYS}, t}\right|=\left|\int_{t_{0}}^{t} \omega_{i\mbox{-}\mathrm{SYS}} d t\right|<\delta_{i\mbox{-}\mathrm{SYS}}^{b o u n d}
\end{equation}
\par From Eq. (\ref{equ22}), because all machines (not matter the machine is a critical machine or a non critical machine) in the entire system are stable, the machines inside Group-CR are certain to maintain stable. Against this background, the EMTR of Machine-CR can be given as:
\begin{equation}
  \label{equ23}
  \delta_{\mathrm{CR}\mbox{-}\mathrm{SYS}}=\frac{\sum_{i \in \Omega_{\mathrm{CR}}} M_{i} \delta_{i\mbox{-}\mathrm{SYS}}}{M_{\mathrm{CR}}}<\frac{\sum_{i \in \Omega_{\mathrm{CR}}} M_{i} \delta_{i\mbox{-}\mathrm{SYS}}^{bound}}{M_{\mathrm{CR}}}
\end{equation}
\par Eq. (\ref{equ23}) indicates that the EMTR is bounded, and thus Machine-CR maintains stable.
\par From the analysis above, one can obtain the following:
\\ (i) Following Eq. (\ref{equ22}), from the individual-machine perspective, the original system is evaluated to maintain stable because all machines in the system are stable, according to the unity principle.
\\ (ii) Following Eq. (\ref{equ23}), from the equivalent-machine perspective, under the dominant group separation pattern, the equivalent system is evaluated to maintain stable because the one-and-only Machine-CR maintains stable.
\par (i) and (ii) fully reveal that the original system maintaining stable is identical to the equivalent system maintaining stable.
\par Taking the simulation case in Sections \ref{section_IIIC} and \ref{section_IIID} as an example. The analysis about the original system stability and the equivalent system stability is given as below.
\vspace*{0.5em}
\\
\textit{Stable case}: In Fig. \ref{fig8}, from the individual-machine perspective, Machines 2 and 3 in the original system maintain stable, and thus the original system maintains stable according to the unity principle.
\par From the equivalent-machine perspective, the dominant $\Omega_{\mathrm{CR}}$ is \{Machine 2, Machine 3\}. The one-and-only equivalent machine maintains stable, and thus the equivalent system also maintains stable.
\\ \textit{Unstable case}: In Fig. \ref{fig10}, from the individual-machine perspective, Machines 2 and 3 in the original system become unstable, and thus the original system becomes unstable according to the unity principle.
\par From the equivalent-machine perspective, the dominant $\Omega_{\mathrm{CR}}$ is \{Machine 2, Machine 3\}. The one-and-only equivalent machine becomes unstable, and thus the equivalent system also becomes unstable.
\vspace*{0.5em}
\par From analysis above, if the trajectory separations occur in the original system, this separation is certain to be reflected in the equivalent system. Therefore, the original system stability is identical to the equivalent system stability.

\subsection{COMPARISON OF THE SYSTEM SEVERITY} \label{section_VIIC}
\noindent \textit{Definition of the original system severity}: The severity depicts margin that the system will become unstable. In brief, if any inner-group motions are slight, all these inner-group machines are stable, and thus the original system can be replaced with the original system. Comparatively, if any inner-group motion is fierce, the inner-group machine instability will occur. Against this background, the severity of the equivalence will become ``lower” than that of the original system, and thus the original system cannot be replaced with the original system.
\par Considering the mechanisms of the inner-group machine motions as analyzed in Section \ref{section_VIIA}, the following two deductions can be obtained
\vspace*{0.5em}
\par \textit{The severity of the equivalent system is close to that of the original system if all the inner-group-machine motions are slight}.
\par \textit{The severity of the equivalent system is lower than that of the original system if any inner-group-machine motion is fierce}.
\vspace*{0.5em}
\par Generally, the inner-group motions are slight in most simulation cases. However, they might also become fierce in some distinctive cases. Under this circumstance, the equivalent machine will show problems in the severity evaluation in TSA. Detailed analysis is given in the following section.

\subsection{A TUTORIAL EXAMPLE ABOUT FIERCE INNER-GROUP MACHINE MOTION} \label{section_VIID}
\begin{figure}[H]
  \centering
  \includegraphics[width=0.4\textwidth,center]{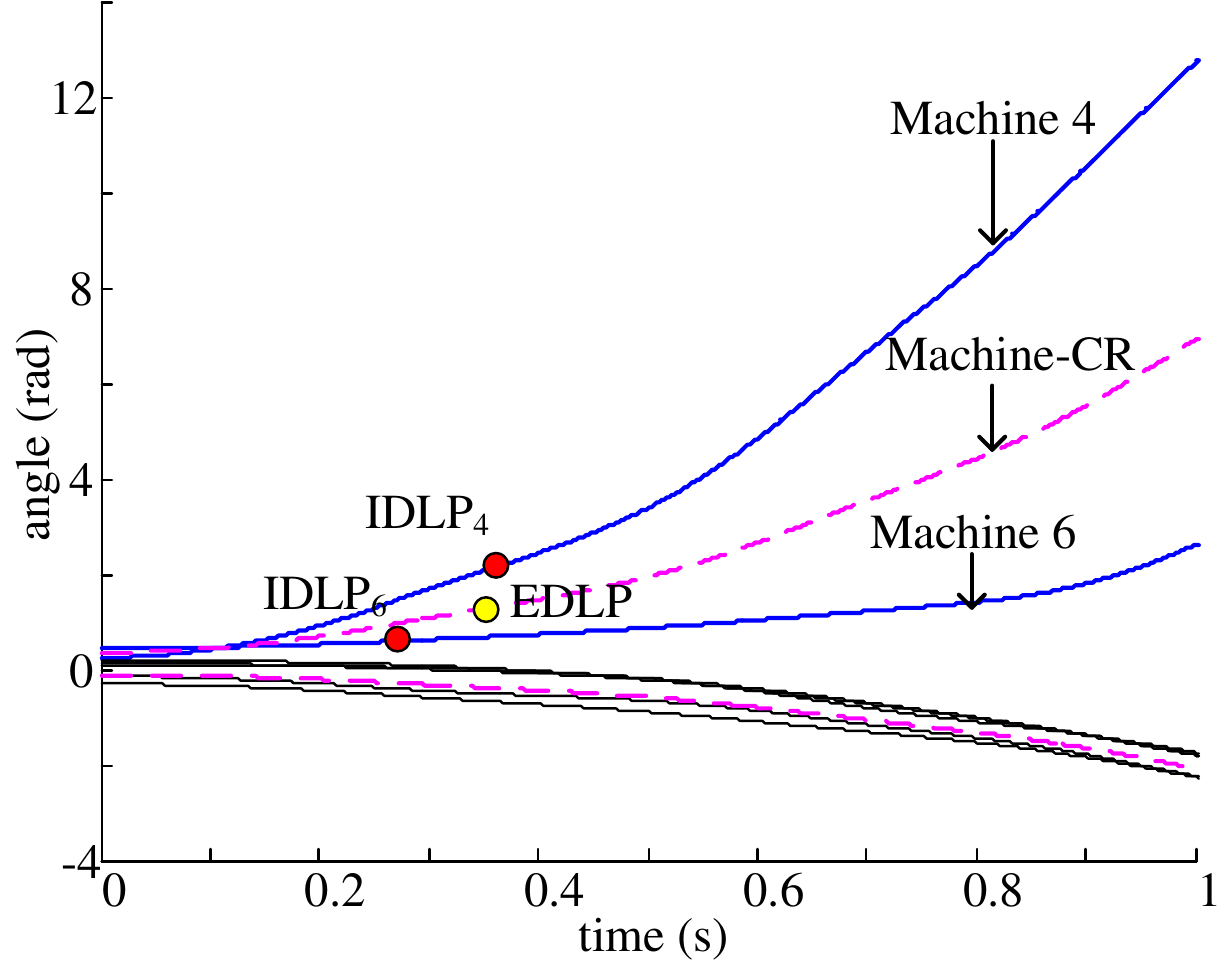}
  \caption{Severe inner-group-machine motion [TS-6, bus-19, 0.260 s].} 
  \label{fig20}  
\end{figure}

\begin{figure}[H]
  \centering
  \includegraphics[width=0.4\textwidth,center]{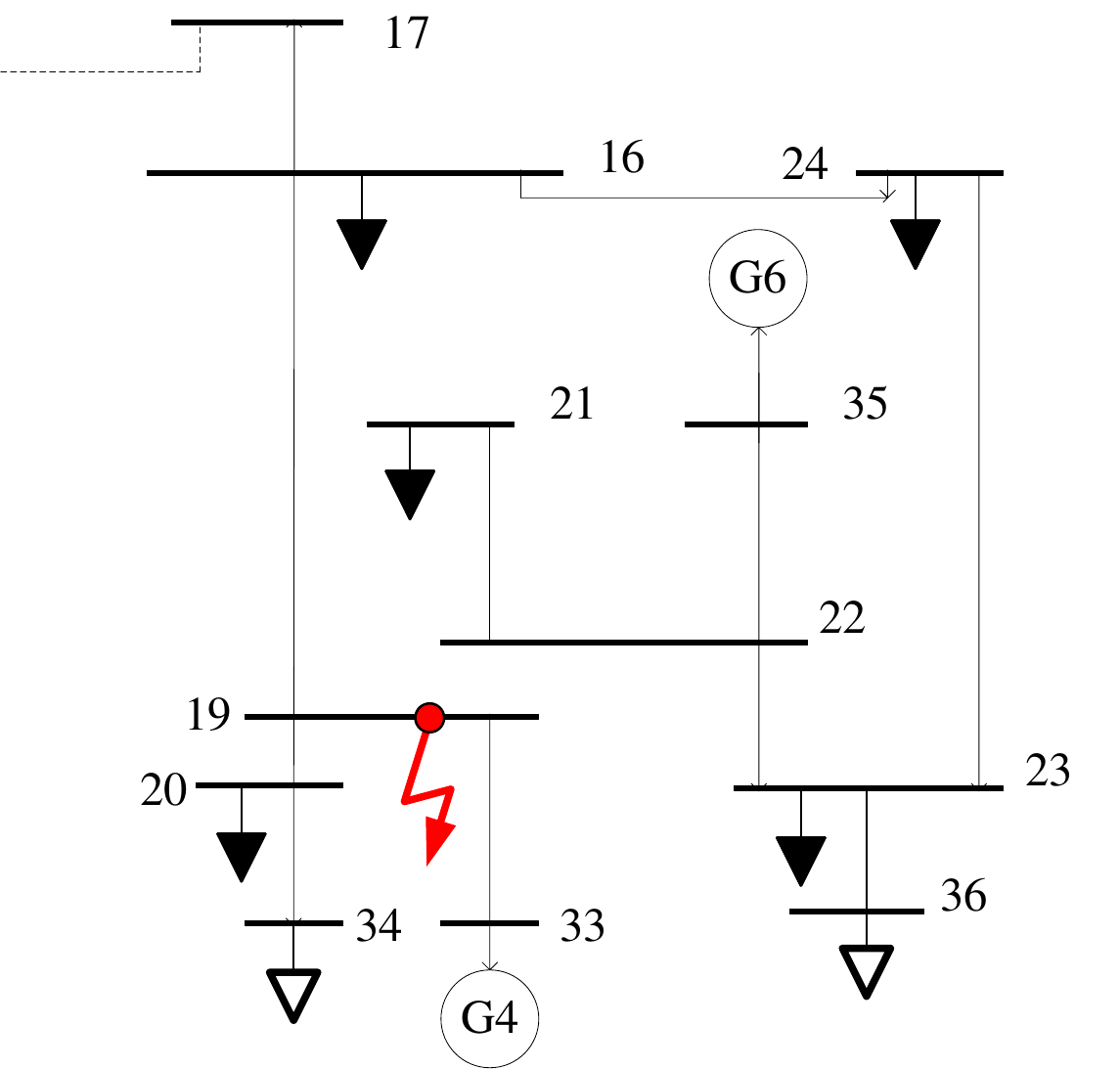}
  \caption{Electric diagram of TS-6.} 
  \label{fig21}  
\end{figure}

The simulation case about the severe inner-group-machine motions is shown in Fig \ref{fig20}. The TS-6 test system is given. TS-6 is based on the modification of TS-1. The two branches L16-21 and L15-16 in TS-1 are disconnected. In addition, Nodes 34, 36 and 38 are modified from PV nodes to PQ nodes.
Part of the electric diagram in TS-6 is shown in Fig. \ref{fig21}. The fault is set as [TS-6, bus-19, 0.260 s]. $\Omega_{\mathrm{CR}}$ is \{Machine 4, Machine 6\}. Notice that the analysis below is based on the mirror system.
\vspace*{-1em}
\vspace*{-0.5em}
\noindent 1) COMPARISION OF THE STABILITY OF THE TWO SYSTEMS
\vspace*{0.5em}
\par From Fig. \ref{fig20}, the EDLP occurs between $\text{IDLP}_6$ and $\text{IDLP}_4$. Therefore, the equivalent system and the original system are characterized as unstable at EDLP and $\text{IDLP}_6$, respectively.
The ``instability” of the equivalent system is characterized later that of the original system. This validates that the stability of the equivalent system is identical to the stability of the original system.
\vspace*{0.5em}
\\ 2) COMPARISION OF THE SEVERITY OF THE TWO SYSTEMS
\vspace*{0.5em}
\par From Fig. \ref{fig20}, the ``severity” of the equivalent system and that of the original system are obtained at EDLP and $\text{IDLP}_4$, respectively. Therefore, the severity of the equivalent system is obtained earlier than that of the original system.
\par Different from the case in Fig. \ref{fig19a} in which all the inner-group-machine motions are slight, the inner-group-machine motions in this case are rather fierce. The inner-group-machine motions inside $\Omega_{\mathrm{CR}}$ are shown in Figs. \ref{fig22} (a) and (b), respectively.
\begin{figure} [H]
  \centering 
  \subfigure[]{%
  \label{fig22a}
    \includegraphics[width=0.4\textwidth]{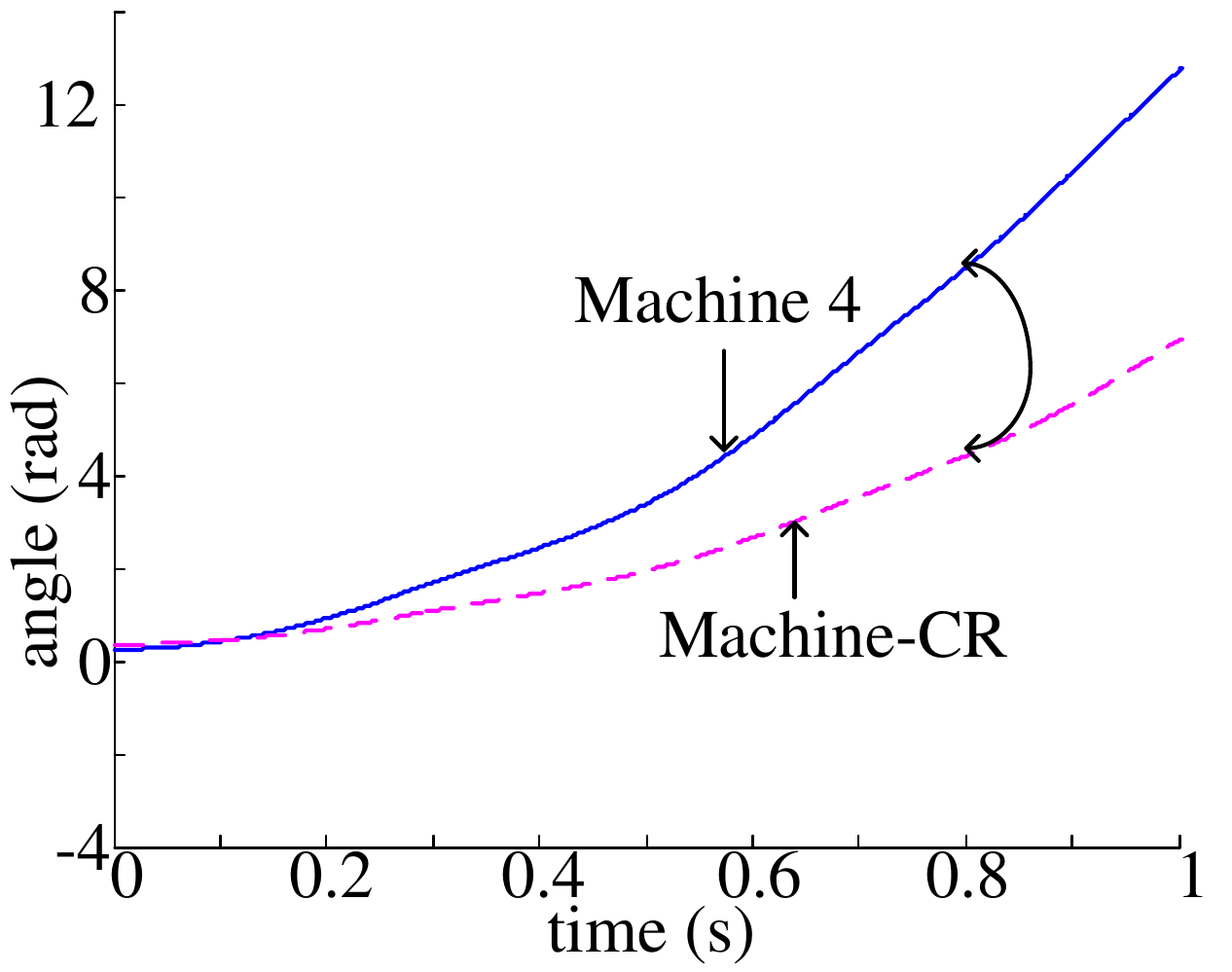}}%
\end{figure} 
\vspace*{-2em}
\addtocounter{figure}{-1}       
\begin{figure} [H]
  \addtocounter{figure}{1}      
  \centering 
  \subfigure[]{%
    \label{fig22b}
    \includegraphics[width=0.4\textwidth]{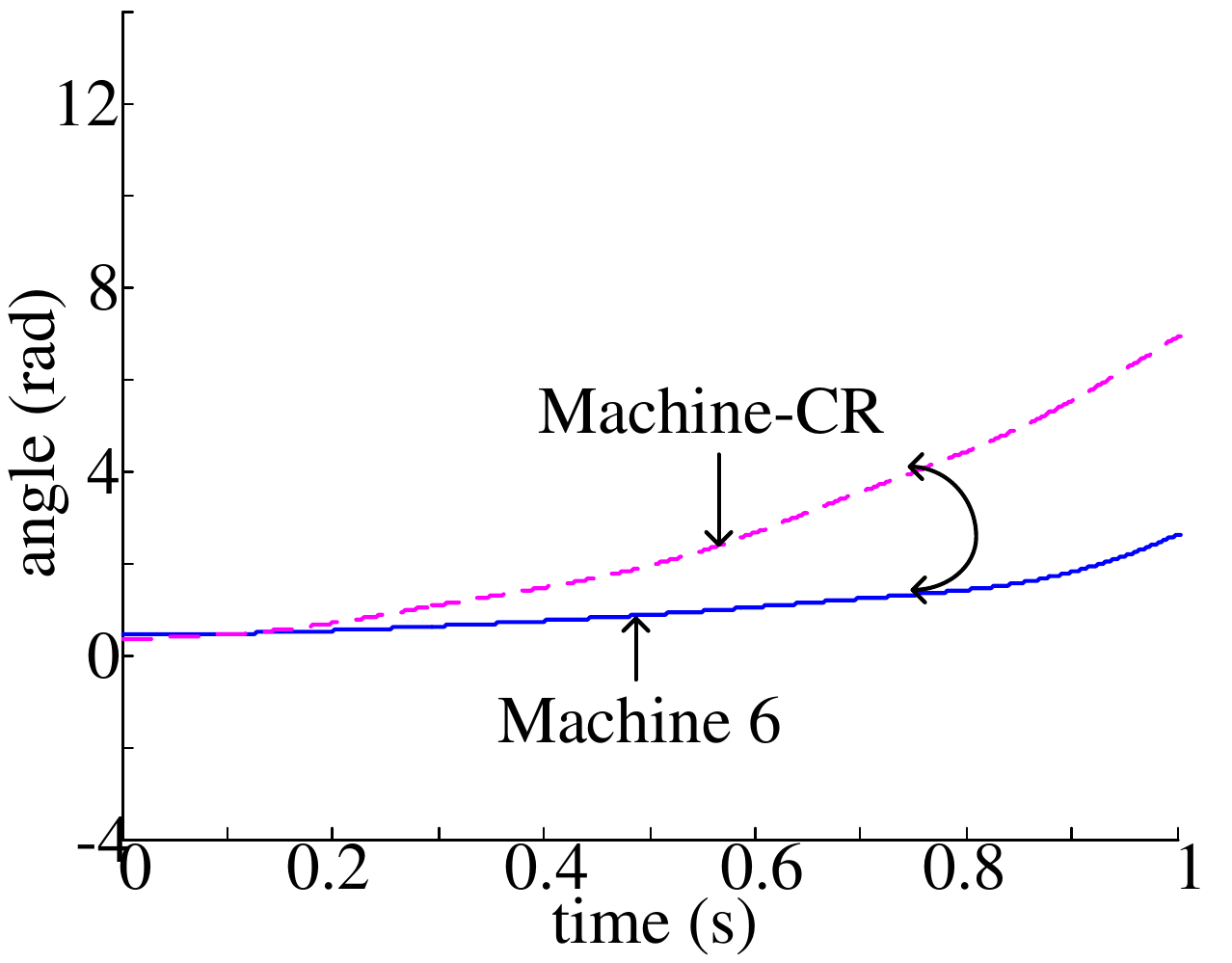}}%
  \caption{Inner-group-machine motions [TS-6, bus-19, 0.260 s]. (a) Motion between Machine 4 and Machine-CR. (b) Motion between Machine 6 and Machine-CR.}%
  \label{fig22}
\end{figure}
\vspace*{-0.5em}
From Fig. \ref{fig22}, the angle difference between Machine 4 and Machine-CR reaches 5.78 rad at 1.000 s. This fully indicates that the ``inner-group-machine instability” already occurs between Machine 4 and Machine-CR.
Under this circumstance, it is quite clear that the severity of the equivalent system is lower than that of the original system. Therefore, because of the inner-group machine instability, the original system cannot be replaced with the equivalent system in this case, even though the stability states of the two systems are completely identical.
\par Detailed analysis about the inner-group machine stability will be given in the companion paper. In the following section, the precise modeling of the mirror system as previously shown in Section \ref{section_IIIA} will be provided.
\\
\section{MIRROR SYSTEM} \label{section_VIII}
\subsection{EQUIVALENT MACHINE IN THE COI-SYS REFERENCE} \label{section_VIIIA}
\noindent \textit{Mirror motion between CR-SYS system and NCR-SYS system:}:The parameters of Machine-CR and Machine-NCR using the COI-SYS reference can be denoted as
\begin{equation}
  \label{equ24}
  \centering
  \begin{aligned}
    &\left\{\begin{array}{l}
        \delta_{\mathrm{CR}\mbox{-}\mathrm{SYS}}=\frac{\sum_{i \in \Omega_{\mathrm{CR}}} M_{i} \delta_{i\mbox{-}\mathrm{SYS}}}{M_{\mathrm{CR}}} \\
        \\
        \omega_{\mathrm{CR}\mbox{-}\mathrm{SYS}}=\frac{\sum_{i \in \Omega_{\mathrm{CR}}} M_{i} \omega_{i\mbox{-}\mathrm{SYS}}}{M_{\mathrm{CR}}} \\
        \\
        f_{\mathrm{CR}\mbox{-}\mathrm{SYS}}=\sum_{i \in \Omega_{\mathrm{CR}}} f_{i\mbox{-}\mathrm{SYS}}
    \end{array}\right.\\
    & \left\{\begin{array}{l}
        \delta_{\mathrm{NCR}\mbox{-}\mathrm{SYS}}=\frac{\sum_{j \in \Omega_{\mathrm{NCR}}} M_{j} \delta_{j\mbox{-}\mathrm{SYS}}}{M_{\mathrm{NCR}}} \\
        \\
        \omega_{\mathrm{NCR}\mbox{-}\mathrm{SYS}}=\frac{\sum_{j \in \Omega_{\mathrm{NCR}}} M_{j} \omega_{j\mbox{-}\mathrm{SYS}}}{M_{\mathrm{NCR}}} \\
        \\
        f_{\mathrm{NCR}\mbox{-}\mathrm{SYS}}=\sum_{j \in \Omega_{\mathrm{NCR}}} f_{j\mbox{-}\mathrm{SYS}}
    \end{array}\right.
  \end{aligned} 
\end{equation}
\par The parameters in Eq. (\ref{equ24}) can also be depicted as
\begin{equation}
  \label{equ25}
  \centering
  \begin{aligned}
    &\left\{\begin{array}{l}
        \delta_{\mathrm{CR}\mbox{-}\mathrm{SYS}}=\delta_{\mathrm{CR}}-\delta_{\mathrm{SYS}} \\
        \\
        \omega_{\mathrm{CR} \mbox{-}\mathrm{SYS}}=\omega_{\mathrm{CR}}-\omega_{\mathrm{SYS}} \\
        \\
        f_{\mathrm{CR}\mbox{-}\mathrm{SYS}}=P_{\mathrm{CR}}-\frac{M_{\mathrm{CR}}}{M_{\mathrm{SYS}}} P_{\mathrm{SYS}}
    \end{array}\right.\\
    & \left\{\begin{array}{l}
        \delta_{\mathrm{NCR}\mbox{-}\mathrm{SYS}}=\delta_{\mathrm{NCR}}-\delta_{\mathrm{SYS}} \\
        \\
        \omega_{\mathrm{NCR}\mbox{-}\mathrm{SYS}}=\omega_{\mathrm{NCR}}-\omega_{\mathrm{SYS}} \\
        \\
        f_{\mathrm{NCR}\mbox{-}\mathrm{SYS}}=P_{\mathrm{NCR}}-\frac{M_{\mathrm{NCR}}}{M_{\mathrm{SYS}}} P_{\mathrm{SYS}}
    \end{array}\right.
  \end{aligned} 
\end{equation}
\par Based on Eq. (\ref{equ25}), The equations of motions of Machine-CR and Machine-NCR in the COI-SYS reference can be depicted as
\begin{equation}
  \label{equ26}
  \centering
  \begin{aligned}
    &\left\{\begin{array}{l}
      { \frac { d \delta _ { \mathrm { CR } \mbox{-} \mathrm { SYS } } } { d t } = \omega _ { \mathrm { CR } \mbox{-} \mathrm { SYS } } } \\
      \\
      { M _ { \mathrm { CR } } \frac { d \omega _ { \mathrm { CR } \mbox{-} \mathrm { SYS } } } { d t } = f _ { \mathrm { CR } \mbox{-}\mathrm{SYS} } }
      \end{array}\right.\\
    & \left\{\begin{array}{l}
      \frac{d \delta_{\mathrm{NCR}\mbox{-}\mathrm{SYS}}}{d t}=\omega_{\mathrm{NCR}\mbox{-}\mathrm{SYS}} \\
      \\
      M_{\mathrm{NCR}} \frac{d \omega_{\mathrm{NCR}\mbox{-}\mathrm{SYS}}}{d t}=f_{\mathrm{NCR}\mbox{-}\mathrm{SYS}}
      \end{array}\right.
  \end{aligned} 
\end{equation}
\par Eq. (\ref{equ26}) indicates that Machine-CR and Machine-NCR in the COI-SYS reference will form CR-SYS system and NCR-SYS system respectively. The formations of the two systems are already shown in Fig. \ref{fig13}.
\par Following Eq. (\ref{equ25}), we have
\begin{equation}
  \label{equ27}
  \left\{\begin{array}{l}
    M_{\mathrm{CR}} \delta_{\mathrm{CR}\mbox{-}\mathrm{SYS}}+M_{\mathrm{NCR}} \delta_{\mathrm{NCR}\mbox{-}\mathrm{SYS}}=0 \\
    \\
    M_{\mathrm{CR}} \omega_{\mathrm{CR}\mbox{-}\mathrm{SYS}}+M_{\mathrm{NCR}} \omega_{\mathrm{NCR}\mbox{-}\mathrm{SYS}}=0 \\
    \\
    f_{\mathrm{CR}\mbox{-}\mathrm{SYS}}+f_{\mathrm{NCR}\mbox{-}\mathrm{SYS}}=0
    \end{array}\right.
\end{equation}
\par Eq. (\ref{equ27}) can be further denoted as
\begin{equation}
  \label{equ28}
  \left\{\begin{array}{l}
    M_{\mathrm{CR}} \frac{d \delta_{\mathrm{CR}\mbox{-}\mathrm{SYS}}}{d t}+M_{\mathrm{NCR}} \frac{d \delta_{\mathrm{NCR}\mbox{-}\mathrm{SYS}}}{d t}=0 \\
    \\
    M_{\mathrm{CR}} \frac{d \omega_{\mathrm{CR}\mbox{-}\mathrm{SYS}}}{d t}+M_{\mathrm{NCR}} \frac{d \omega_{\mathrm{NCR}\mbox{-}\mathrm{SYS}}}{d t}=0
    \end{array}\right.
\end{equation}
\par Eq. (\ref{equ28}) reflects that the ``mirror motion” can be found between Machine-CR and Machine-NCR. In particular, in the COI-SYS reference, Machine-NCR moves in the opposite direction of Machine-CR, and this motion can also be seen as a constant ratio of the motion of Machine-CR. In addition, according to Eq. (\ref{equ27}), the two systems will become unstable simultaneously ($f_{\mathrm{CR}\mbox{-}\mathrm{SYS}}=f_{\mathrm{NCR}\mbox{-}\mathrm{SYS}}=0$).
\par
\begin{figure}[H]
  \centering
  \includegraphics[width=0.45\textwidth,center]{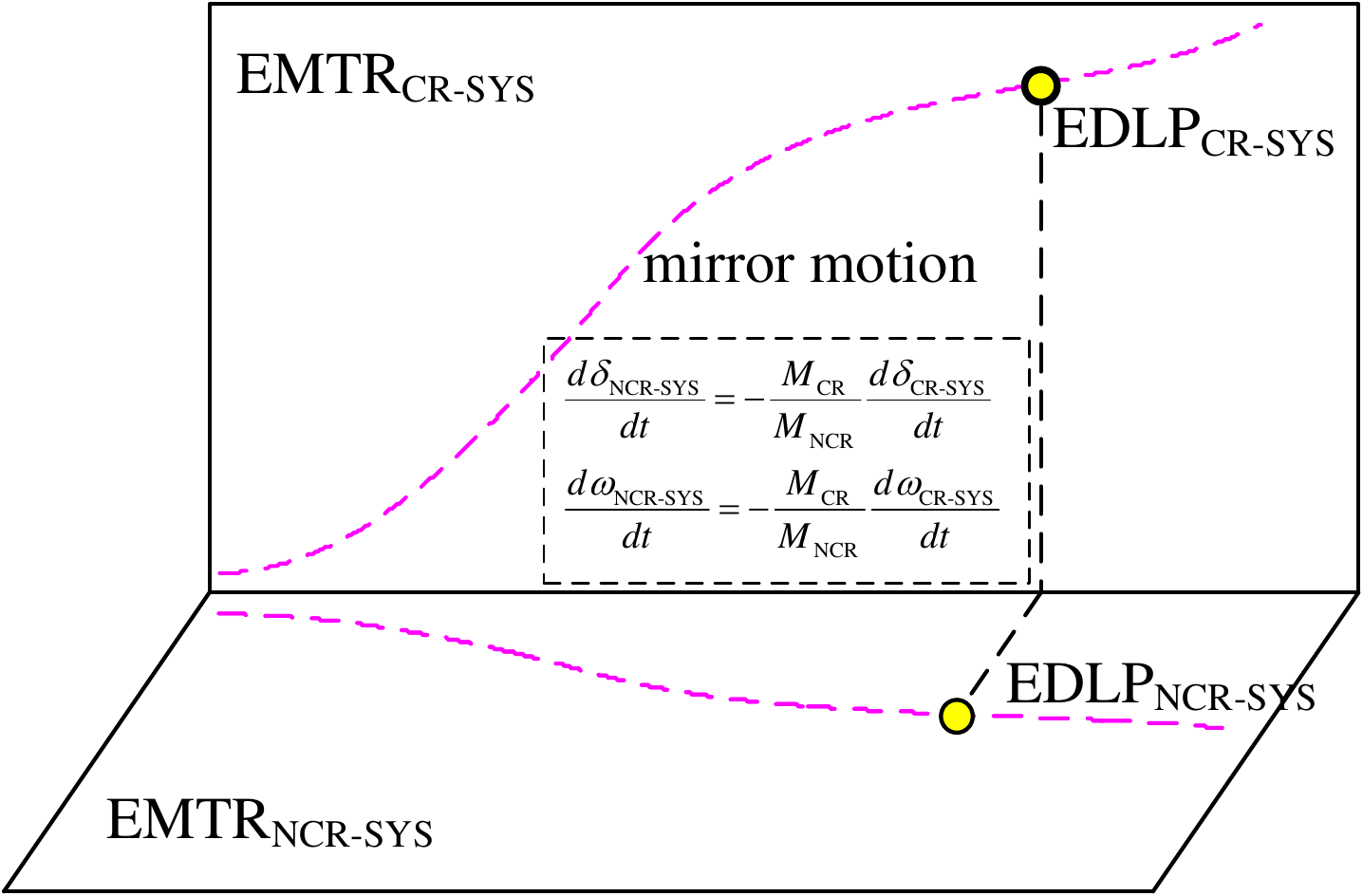}
  \caption{Mirror motion between CR-SYS system and NCR-SYS system.} 
  \label{fig23}  
\end{figure}

 Demonstration about the mirror motion is shown in Fig. \ref{fig23}.
\noindent\textit{Mirror motion between CR-SYS system and CR-NCR system}: We further extend the analysis above in to the CR-NCR system case.
\par The parameters of the CR-NCR system that are re-depicted in the COI-SYS reference are denoted as
\begin{equation}
  \label{equ29}
  \left\{\begin{array}{l}
    \delta_{\mathrm{CR}\mbox{-}\mathrm{NCR}}=\delta_{\mathrm{CR}\mbox{-}\mathrm{SYS}}-\delta_{\mathrm{NCR}\mbox{-}\mathrm{SYS}} \\
    \\
    \omega_{\mathrm{CR}\mbox{-}\mathrm{NCR}}=\omega_{\mathrm{CR}\mbox{-}\mathrm{SYS}}-\omega_{\mathrm{NCR}\mbox{-}\mathrm{SYS}} \\
    \\
    f_{\mathrm{CR}\mbox{-}\mathrm{NCR}}=f_{\mathrm{CR}\mbox{-}\mathrm{SYS}}-\frac{M_{\mathrm{CR}}}{M_{\mathrm{NCR}}} f_{\mathrm{NCR}\mbox{-}\mathrm{SYS}}
    \end{array}\right.
\end{equation}
\par Based on Eqs. (\ref{equ27}) and (\ref{equ29}), we have
\begin{equation}
  \label{equ30}
  \centering
  \left\{\begin{array}{l}
    M_{\mathrm{NCR}} \delta_{\mathrm{CR}\mbox{-}\mathrm{NCR}}=M_{\mathrm{SYS}} \delta_{\mathrm{CR}\mbox{-}\mathrm{SYS}} \\
    \\
    M_{\mathrm{NCR}} \omega_{\mathrm{CR}\mbox{-}\mathrm{NCR}}=M_{\mathrm{SYS}} \omega_{\mathrm{CR}\mbox{-}\mathrm{SYS}} \\
    \\
    M_{\mathrm{NCR}} f_{\mathrm{CR}\mbox{-}\mathrm{NCR}}=M_{\mathrm{SYS}} f_{\mathrm{CR}\mbox{-}\mathrm{SYS}}
    \end{array}\right.
\end{equation}
\par From Eq. (30), one can obtain
\begin{equation}
  \label{equ31}
  \left\{\begin{array}{l}
    M_{\mathrm{NCR}} \frac{d \delta_{\mathrm{CR}\mbox{-}\mathrm{NCR}}}{d t}-M_{\mathrm{SYS}} \frac{d \delta_{\mathrm{CR}\mbox{-}\mathrm{SYS}}}{d t}=0 \\
    \\
    M_{\mathrm{NCR}} \frac{d \omega_{\mathrm{CR}\mbox{-}\mathrm{NCR}}}{d t}-M_{\mathrm{SYS}} \frac{d \omega_{\mathrm{CR}\mbox{-}\mathrm{SYS}}}{d t}=0
    \end{array}\right.
\end{equation}{
\par Eq. (\ref{equ31}) reflects that the ``mirror motion” can also be found between CR-SYS system and the CR-NCR system. The two systems will also become unstable simultaneously along time horizon ($f_{\mathrm{CR}\mbox{-}\mathrm{NCR}}=f_{\mathrm{CR}\mbox{-}\mathrm{SYS}}=0$).
In addition, one can simply prove that the stability margin of the CR-SYS system is identical to that of the CR-NCR system.
\par The analysis above validates all the characteristics of the mirror system as given in Section \ref{section_IIIA}.

\section{CONCLUSIONS} \label{section_IX}
In this paper the mechanism of the equivalent-machine is analyzed. The equivalent machine is established based on the ``motion equivalence” rather than the ``energy superimposition” of all machines in the system.
The equivalent-machine trajectories are first established based on the group separation. Then, the relative motion of the two equivalent-machine trajectories is modeled through the CR-NCR system with strict NEC and EAC characteristic.
Against this background, the equivalent machine becomes the ``one-and-only” machine in the equivalent system under the dominant group separation pattern. The transient characteristics of the equivalent machine are explained from the individual-machine perspective. It is clarified that the equivalent machine strictly follows all the machine paradigms.
These strict followings bring the two advantages in TSA. The transient stability concepts are also defined in an equivalent-machine manner. In particular, the one-and-only equivalent-machine swing is clearly depicted through the EMPP (trajectory-depiction advantage), and the critical stability of the equivalent system is decided through the critical stability of the one-and-only equivalent machine (the two advantages).
In the end of the paper, it is clarified that the equivalent-system stability and the original-system stability are completely identical. However, the original-system severity might be different from the equivalent-system severity under distinctive circumstances once the inner-group machine motions become fierce.
\par In the following paper, the inner-group-machine stability will be analyzed based on the machine paradigms. This may essentially reveal the complicated relationships between the original system and the equivalent system.

%

%
%
%




\end{document}